\documentclass[12pt,epsf,amssymb]{article}

\usepackage{graphicx}
\usepackage{amsfonts}
\usepackage{amsmath}
\usepackage[usenames]{color}
\usepackage{amssymb}
\usepackage[mathscr]{eucal} 
\usepackage{url}

\def\Z{\mathbb{Z}}
\def\Q{\mathbb{Q}}

\def\P{\mathbb{P}}

\def\Im{\mathrm{Im}}
\def\Re{\mathrm{Re}}

\begin{document}

\baselineskip 0.6cm
\newcommand{\vev}[1]{ \left\langle {#1} \right\rangle }
\newcommand{\bra}[1]{ \langle {#1} | }
\newcommand{\ket}[1]{ | {#1} \rangle }
\newcommand{\Dsl}{\mbox{\ooalign{\hfil/\hfil\crcr$D$}}}
\newcommand{\nequiv}{\mbox{\ooalign{\hfil/\hfil\crcr$\equiv$}}}
\newcommand{\nsupset}{\mbox{\ooalign{\hfil/\hfil\crcr$\supset$}}}
\newcommand{\nni}{\mbox{\ooalign{\hfil/\hfil\crcr$\ni$}}}
\newcommand{\nin}{\mbox{\ooalign{\hfil/\hfil\crcr$\in$}}}
\newcommand{\Slash}[1]{{\ooalign{\hfil/\hfil\crcr$#1$}}}
\newcommand{\EV}{ {\rm eV} }
\newcommand{\KEV}{ {\rm keV} }
\newcommand{\MEV}{ {\rm MeV} }
\newcommand{\GEV}{ {\rm GeV} }
\newcommand{\TEV}{ {\rm TeV} }

\newcommand{\hh}[1]{{\color{green}  #1}}

\def\diag{\mathop{\rm diag}\nolimits}
\def\tr{\mathop{\rm tr}}

\def\Spin{\mathop{\rm Spin}}
\def\SO{\mathop{\rm SO}}
\def\O{\mathop{\rm O}}
\def\SU{\mathop{\rm SU}}
\def\U{\mathop{\rm U}}
\def\Sp{\mathop{\rm Sp}}
\def\SL{\mathop{\rm SL}}

\def\change#1#2{{\color{blue}#1}{\color{red} [#2]}\color{black}\hbox{}}

\begin{titlepage}
 
\begin{flushright}
IFT-UAM/CSIC-14-104 \\
IPMU14-0321
\end{flushright}
 
\vskip 1cm
\begin{center}
 
{\large \bf Issues in Complex Structure Moduli Inflation}
 
\vskip 1.2cm
 
Hirotaka Hayashi$^1$, Ryo Matsuda$^2$ and Taizan Watari$^2$

 \vskip 0.4cm
 
 {\it $^1$Instituto de F\'{\i}sica Te\'orica UAM/CSIC, Cantoblanco, 28049 Madrid, Spain
\\[2mm]

   $^3$Kavli Institute for the Physics and Mathematics of the Universe, 
  University of Tokyo, Kashiwa-no-ha 5-1-5, 277-8583, Japan
  }
  \vskip 1.5cm
  
\abstract{
Supersymmetric compactification with moderately large radius 
(${\rm Re}\vev{T} \sim {\cal O}(10)$ or more) not only accommodates 
supersymmetric unification, but also provides candidates for an inflaton 
in the form of geometric moduli; the value of ${\rm Re}\vev{T} > 1$ may be 
used as a parameter that brings corrections to the inflaton potential under 
control. Motivated by a bottom-up idea ``right-handed sneutrino inflation'' 
scenario, we study whether complex structure moduli can play some role 
during the slow-roll inflation and/or reheating process in this moderately 
large radius regime.
Even when we allow a tuning introduced by Kallosh and Linde, 
the barrier of volume stabilization potential from gaugino condensation 
racetrack superpotential can hardly be as high as $(10^{16} \; \GEV)^4$ 
for generic choice of parameters in this regime. It is also found 
that even very small deformation of complex structure during 
inflation/reheating distorts the volume stabilization potential, so that 
the volume stabilization imposes tight constraints on large-field inflation 
scenario involving evolution of complex structure moduli. 
A few ideas of satisfying those constraints in string theory are also 
discussed.  } 
 
\end{center}
\end{titlepage}
 
 
\section{Introduction}
\label{sec:intro}

Inflation provides an ideal opportunity to catch a glimpse of 
physics at very high energy. Time-evolution of an inflaton is sensitive 
to Planck-scale suppressed corrections, in small field inflation models 
and large field models alike; although the energy density during inflation 
was not as high as $\rho \sim M_{\rm Pl}^4$, where 
$M_{\rm Pl} \simeq 2.43 \times 10^{18} \GEV$, we still need to have theoretical 
control over coherent deformation of the universe (inflaton field value) 
that is more than ${\cal O}(1)$ in $M_{\rm Pl}$ unit. String theory being 
virtually the only known candidate for calculable quantum theory of gravity, 
it is worthwhile to study what the ``control'' is like, and what the 
microscopic picture of the ``coherent deformation of the universe'' is 
like, in string theory. 

String theory has another advantage of being an ``all-in-one package'';  
one cannot choose a set of compactification manifold and brane configuration 
for inflation and another for physics of quarks and leptons. 
This is in sharp contrast against the traditional framework of low-energy 
effective field theory. Models for inflaton and those for 
our particles (quarks, leptons, photons etc.) can be discussed 
separately in low-energy effective field theory, and a set-up necessary 
for successful inflation rarely imposes constraints on the models of 
elementary particles, or vice versa. In string theory, however, 
we need to have both the inflaton and the particle sector together 
in a common package of compactification manifold and brane configuration, 
and the post inflationary evolution of the universe\footnote{Especially 
it should have a graceful exit from inflation and subsequent reheating 
to our particle sector, but not too much to other sectors (incl. 
gravitino, axino, (s)axion, hidden sector particles and moduli fields).} 
needs to be worked out in the common package. 
This is highly a non-trivial study, 
but such a tight constraint from string theory can be exploited to 
uncover the history of the universe in the early stage. 

In this article, we are motivated by a bottom-up idea that the inflaton 
may be identified with the supersymmetry partner of right-handed neutrinos 
(right-handed sneutrino inflation scenario) \cite{sneutrino}. 
Post-inflationary thermal 
history has been studied very well, and it is known that this scenario 
passes various constraints from phenomenology. This scenario comes with 
some assumptions on K\"{a}hler potential, however, and it is a legitimate 
question whether this bottom-up idea can be accommodated in some string 
compactification or not. Right-handed neutrino chiral multiplets are 
known to be identified with a part of complex structure moduli in F-theory 
compactifications in the matter parity 
scenario \cite{RHnu-Ftheory},\footnote{This is equivalent,  
through string duality, to a statement that vector bundle moduli are 
identified with right-handed neutrino chiral multiplets in Heterotic string 
compactifications \cite{Witten-SU(3)}.} and this observation motivates 
us to study what it takes for complex structure moduli to drive (or at least to be relevant to) inflation
in Type IIB/F-theory compactifications in general. 

Mirror symmetry indicates that the K\"{a}hler potential of some kind of 
complex structure moduli has properties that we expect for that of the 
K\"{a}hler moduli, as pointed out also recently in the context of 
inflation \cite{Silverstein} (see also \cite{Hebecker,cpxstrinflation}).\footnote{Recently, there have been also progress in large field inflation scenarios using other moduli in string compactification \cite{inflation}.}
Since the K\"{a}hler potential of K\"{a}hler moduli has an approximate 
shift symmetry in the regime supergravity approximation is valid, 
the K\"{a}hler potential of complex structure moduli should also have 
an approximate shift symmetry near the large complex structure limit. 
When flux is introduced in Type IIB/F-theory compactifications, 
complex structure moduli dependent superpotential is also 
generated \cite{GVW}. The Majorana mass of right-handed neutrinos, which 
is tied with the mass of the complex structure moduli, sets a scale 
somewhat below the Kaluza--Klein scale \cite{KST, RHnu-Ftheory}, which 
is a necessary ingredient of inflation \cite{sneutrino}. 
A natural question is, then, whether this set-up can be relevant, in any way, 
to the slow-roll inflation process. We address this question in this article, 
and intend to share our thoughts with experts around the world.

In section \ref{sec:KL-problem}, we combine the discussion of 
Refs. \cite{KL-problem} and \cite{GKP} 
and arrive at two observations; i) the stabilization of K\"{a}hler moduli 
during inflation sets a stringent constraint on the evolution of complex 
structure moduli during inflation, and ii) mass-eigenstates at the vacuum 
reached after inflation are generically mixtures of K\"{a}hler moduli and 
complex structure moduli fields. In order to work out the constraint in the 
observation (i) in detail, one needs to estimate the prefactor in the gaugino 
condensation superpotential, first of all; therefore we refine the existing 
estimate of the prefactor in section \ref{sec:scale}. In section \ref{sec:barrier}, we derive 
constraints on how much the field value of complex structure moduli during inflation/reheating can 
be different from the vacuum value,  provided 
the K\"{a}hler moduli field is stabilized by the racetrack superpotential 
from two gaugino condensates; we employed a phenomenological approach 
so that the constraint can be stated without referring to the choice 
of a Calabi--Yau 3-fold for Type IIB compactification. 
In sections \ref{ssec:first} and \ref{ssec:alg-nmb}, such a 
phenomenological and robust constraint is translated into the language 
of Type IIB flux compactification superpotential, and studied 
in combination with how to realize the tuning of Kallosh--Linde for 
supersymmetric volume stabilization \cite{KL-problem}. We see that the constraint is so 
stringent that very little e-fold can be earned generically by evolution of the complex 
structure moduli fields. In section \ref{ssec:solution}, we discuss a few 
possibilities of making the right-handed sneutrino inflation scenario and/or 
complex structure moduli driven inflation still at work, while satisfying the constraint. 

\section{Interplay between K\"{a}hler and Complex Structure Moduli under 
the Kallosh--Linde Tuning}
\label{sec:KL-problem}

\subsection{Volume Stabilization and Large Field/High Scale Inflation}

One of the most important lessons from studies on D-brane inflation almost 
a decade ago was that stabilization of the K\"{a}hler moduli of a 
compactification is essential \cite{KKLMMT, KL-problem} (see also \cite{Buch}). 
This remains true in the current situation. 

To get started, let us remind ourselves that the scalar potential for the complex 
structure moduli in flux compactification was shown to be positive definite 
in \cite{GKP} under two assumptions. The first assumption is that 
the superpotential of the 4D effective theory is independent 
of K\"{a}hler moduli, $\partial_T W = 0$, and the other is that the 
K\"{a}hler potential of the K\"{a}hler moduli $T$ is of no-scale type:
\begin{equation}
 \frac{K}{M_{\rm Pl}^2} = - 3 \ln [{\rm Re}(T)] + {\cal K}^{(\rm cpx)}(z,z^\dagger). 
\label{eq:Kahler-noscale+cpx}
\end{equation}
Then the superpotential depends only on complex structure moduli $z$, 
$W = W^{(\rm cpx)}(z)$ (in the context of Type IIB Calabi--Yau orientifolds, 
$z$ consists of both the axion-dilaton chiral multiplet $\tau$ and complex 
structure moduli $\zeta$ of a 3-fold).\footnote{To readers unfamiliar with 
string theory: it will be helpful to read a brief review in 
section \ref{sssec:review-m.quintic}, in order to get the feeling of what 
we have in mind for ${\cal K}^{({\rm cpx})}(z,z^\dagger)$ and $W^{({\rm cpx})}(z)$.}
The positive-definiteness of the potential is simply due to cancellation 
within the second term of 
\begin{equation}
 V = e^{K/M_{\rm Pl}^2} K^{\bar{z}z} |D_zW|^2 + 
  e^{K/M_{\rm Pl}^2} \left[ K^{\bar{T}T} |D_TW|^2
                        - 3 \left| \frac{W}{M_{\rm Pl}} \right|^2 
                \right],
\label{eq:scalar-potential-ZandT}
\end{equation}
where 
\begin{equation}
K^{\bar{T}T} M_{\rm Pl}^2 = \frac{(T+\bar{T})^2}{3}, \qquad 
\frac{K_T}{M_{\rm Pl}^2} = - \frac{3}{(T+\bar{T})};  
\end{equation}
the scalar potential $V$ is given by the positive definite first term 
after the cancellation.

The K\"{a}hler moduli needs to be stabilized, however.  
Since the $T$-independence of the superpotential is a crucial ingredient 
of the assumptions leading to the positive-definiteness, one should not 
expect this positive-definiteness to hold true, even after non-perturbative 
effects generate $T$-dependent terms in the effective superpotential. 
Suppose for now\footnote{\label{fn:multi-h11}
It looks as if we assume that there is only 
one K\"{a}hler modulus here, while it is well-known that the orientifold-even 
part of $h^{1,1}$ of a Calabi--Yau 3-fold for Type IIB compactification 
($h^{1,1}$ of Calabi--Yau 4-fold for F-theory) is not always 1. Even in 
cases with $h^{1,1}_+ >1$ ($h^{1,1}>1$ resp.), however, it is possible that 
there is only one independent K\"{a}hler modulus field after flux is 
introduced, because the Fayet--Iliopoulos parameter $\xi$ of a U(1) gauge 
field on a 7-brane is of the form $\xi \sim \int \omega \wedge F$ for a 
non-trivial flux $F$ on the 7-brane world-volume \cite{JL}. It is still 
an assumption in such 
cases that the effective K\"{a}hler potential of such an independent 
K\"{a}hler modulus remains to be in the form of (\ref{eq:Kahler-noscale+cpx}).
\label{footnote3}} 
that the K\"{a}hler potential remains to be the one 
in (\ref{eq:Kahler-noscale+cpx}), and the effective superpotential is 
of the form 
\begin{equation}
 W^{({\rm tot})} = W^{(T)}(T,z) + W^{(\rm cpx)}(z).
\label{eq:super-Tz-Z}
\end{equation}
There remains partial cancellation in the second term 
of (\ref{eq:scalar-potential-ZandT}), but the cancellation is no longer 
complete:
\begin{equation}
 V = e^{K/M_{\rm Pl}^2} \left[ K^{\bar{z}z} |D_z W|^2 + K^{\bar{T}T} |\partial_T W|^2
     + \frac{K^{\bar{T}T}}{M_{\rm Pl}^2} \left\{ 
           (\partial_T W) K_{\bar{T}} \overline{W} + {\rm h.c.} \right\} 
    \right].
 \label{eq:scalar-potential-ZandT-partialNoScaleCancl}
\end{equation}
The last term containing $\left\{ (\partial_T W) K_{\bar{T}} \overline{W} 
+ {\rm h.c.}\right\}$ can be either positive or negative, and the scalar 
potential $V$ is not guaranteed to be positive 
definite. 

When the coherent value of the effective superpotential $W$ 
during inflation is too much different from the vacuum value, the potential 
for K\"{a}hler moduli $T$ stabilization is deformed so much that the internal 
space may start to decompactify. The condition of volume stabilization 
therefore has to be imposed on any kinds of complex structure moduli inflation models
(particularly large field/high scale inflation models). We will elaborate more on this issue 
in later sections. 

In order to support inflation in the complex structure moduli field 
with a high energy scale involved, it is better to stabilize the volume 
at as high energy scale as possible; the original motivation of the 
right-handed sneutrino inflation scenario \cite{sneutrino} was to implement 
chaotic inflation indeed.  To stabilize the K\"{a}hler moduli 
at supersymmetric level, we need a fine tuning \cite{KL-problem}, and 
indeed only one fine tuning is necessary. This is to assume that  
\begin{equation}
 \langle W^{(T)}(\vev{T},\vev{z}) + W^{(\rm cpx)}(\vev{z}) \rangle = 0
\label{eq:cond-KL-min}
\end{equation}
at the minimum (well after the last inflation), where this condition is 
imposed for the value of $z = \vev{z}$ and $T = \vev{T}$ determined by 
\begin{equation}
  \langle \partial_T W^{(T)}(T,z) \rangle = 0, \quad 
 \langle \partial_z (W^{(T)}(T,z) + W^{(\rm cpx)}(z)) \rangle = 0.
 \label{vevTz} 
\end{equation}
With these conditions, $z = \vev{z}$ and $T = \vev{T}$ give a minimum of $V$ 
with vanishing cosmological constant (in the tree-level approximation of 
4D supergravity).\footnote{Supersymmetry breaking and non-vanishing 
cosmological constant are ignored as subleading effects throughout 
this article.} We refer to the condition (\ref{eq:cond-KL-min}) as 
Kallosh-Linde tuning in this article. 

For small fluctuations around this supersymmetric minimum, where the 
vacuum energy also vanishes, the scalar potential is positive definite.
When the field value differs too much from their vacuum value, however, 
the scalar potential is no longer guaranteed to be positive, as we 
have already discussed 
below (\ref{eq:scalar-potential-ZandT-partialNoScaleCancl}).
The question is how much deformation in the moduli fields are allowed, 
and this is what we study in the following sections. 

\subsection{Comments on Reheating Process }

One can also work out the mass matrix around such a minimum,\footnote{
Note that the following discussion as well as that in the next section 
remains valid even when the K\"{a}hler potential is not precisely the 
same as the no-scale type $\propto - 3 \ln [{\rm Te}(T)]$, or 
there is more than one independent K\"{a}hler moduli chiral multiplet.
No matter how many independent K\"{a}hler moduli are left after taking account 
of the D-term from fluxes on D-branes (see footnote~\ref{footnote3}), their masses can be much larger 
than the gravitino mass as long as the tuning (\ref{eq:cond-KL-min}) 
is achieved somehow \cite{Krasnikov}.\label{footnote:kahlermass}} which is 
relevant information on the reheating process. The quadratic part 
of the action at the minimum is in the form of 
\begin{equation}
 {\cal L}^{(2)} =
  - \vev{K_{T\bar{T}}} |\partial (\delta T)|^2
  - \vev{K_{z\bar{z}}} |\partial (\delta z)|^2 
  - \left( (\delta T)^\dagger, (\delta z)^\dagger \right) \cdot M \cdot 
    \left( \begin{array}{c} \delta T \\ \delta z \end{array} \right), 
\end{equation}
where $\delta T =T-\vev{T}$, $\delta z =z-\vev{z}$ and the mass matrix is given by 
\begin{eqnarray}
  M & = & e^{K/M_{\rm Pl}^2} K^{\bar{T}T} 
   \left( \begin{array}{cc}
       |\partial^2_T W|^2 & 
       \overline{(\partial_T^2 W)}(\partial_{z}\partial_T W) \\
       \overline{(\partial_{z} \partial_T W)}(\partial_T^2 W) & 
       |\partial_z \partial_T W|^2 \end{array} \right) \nonumber \\
&&  + e^{K/M_{\rm Pl}^2} K^{\bar{z}z} 
   \left( \begin{array}{cc}
       |\partial_T \partial_z W|^2 & 
       \overline{(\partial_{T} \partial_z W)}(\partial_z^2 W) \\
       \overline{(\partial_z^2 W)} (\partial_{T} \partial_z W) & 
       | \partial_z^2 W |^2  \end{array} \right)
\label{eq:phys-mass}
\end{eqnarray}
evaluated at the minimum.
The physical mass matrix $m^2$ is obtained by rescaling the fluctuations 
$\delta T$ and $\delta z$ so that the kinetic terms become canonical at 
the minimum (i.e., by sandwiching $M$ by 
$\diag(\vev{K^{\bar{T}T}}, \vev{K^{\bar{z}z}})^{1/2}$). 
Since it is common place in string compactifications\footnote{Footnote~\ref{fn:threshold} provides a pedagogical explanation for this.} that the term $W^{(T)}(T,z)$ depends on the moduli $z$, 
there are mixing terms in the mass matrix. 
All the mass eigenstates around the minimum therefore possess interactions 
of both $\delta T$ and $\delta z$, when all the relevant mass scales are high 
(as assumed in the context of tuning (\ref{eq:cond-KL-min})). If the 
fluctuation $\delta z$ has a renormalizable coupling with particles 
in the Standard Model (remember that the right-handed neutrinos do have one), 
this means that all the energy of coherent oscillation in $\delta T$ and 
$\delta z$ is converted quickly into that of radiation in the Standard Model, 
and the cosmological moduli problem \cite{moduli-problem} is 
avoided;\footnote{The cosmological moduli problem is avoided also by the large moduli mass (see footnote~\ref{footnote:kahlermass}).}  
matter parity is ignored for simplicity of the argument for now, and 
we will come to this point shortly.   

When the physical masses $m_{T\bar{T}}^2$ and $m_{z\bar{z}}^2$ are comparable, 
the $\delta T$--$\delta z$ mixing due to the $z$-dependence of $W^{(T)}(T,z)$
can be sizable, as this will become clearer after the discussion in section 
\ref{sec:scale}.
It is often perceived that the interactions between K\"{a}hler moduli 
and complex structure moduli are suppressed by Planck scale (or string scale 
$M_s$); that is certainly true, but in the cases where the relevant energy 
scale of gaugino condensation (K\"{a}hler moduli stabilization) is very 
high (as required in high scale inflation), mass mixing 
$W \supset m \; (\delta T) \cdot (\delta z)$ comes at the order of 
$m \approx ({\rm high~scale})^2/M_s$, and the renormalizable interactions  
such as $W \supset \lambda (\delta T)(\delta z)^2$
have coefficients $\lambda \approx ({\rm high~scale})/M_s$.  
Therefore, decay processes through this mixing will have rates of order 
$\Gamma \approx ({\rm high~scale})^3/M_s^2$, which is often sizable in the 
context of thermal history of the universe after inflation. 

In a more realistic set-up of supersymmetric compactification, 
one might wish to introduce matter parity for proton stability.
Such a scenario assumes that a $\Z_2$ symmetry is restored at the minimum 
$z = \vev{z}$ and $T = \vev{T}$, so that all the fluctuations are classified 
into the $\Z_2$-even and $\Z_2$-odd sectors; let $\hat{\Phi}_+$ and 
$\hat{\Phi}_-$ denote such even and odd mass eigenstates, respectively. 
The mass/kinetic mixing as above is found only within these two sectors 
separately. Because there can be interactions such as 
$W \supset \hat{\Phi}_+ \hat{\Phi}_- \hat{\Phi}_-$, various chains of preheating and (on-shell or virtual) 
cascade decay processes may be at work, which means that energy can be 
transferred from the $\Z_2$-even sector to the odd sector, or vice versa. 
Thus realistic story of reheating process would not be as simple as in the 
discussion above without matter parity. 
How much fraction of coherent oscillation energy is in the $\Z_2$-even 
sector of $(\delta T, \delta z)$ at the end of inflation depends on details 
of inflation models. The preheating and decay processes depend very much 
of the spectrum of the moduli fields at the vacuum. It also makes a 
big difference in the thermal history whether there are $\Z_2$-even moduli 
fields with renormalizable couplings with the Standard Model particles 
($W \supset \hat{\Phi}_+ \cdot H_u \cdot H_d$ or 
$\supset \hat{\Phi}_+ \cdot {\bf 5} \cdot \bar{\bf 5}$ 
(vector-like SUSY-breaking messengers)). 
For all these model-dependence, we do not try to develop discussion on the 
reheating process further than this in this article. 

\section{Setting Scales}
\label{sec:scale}

In flux compactification of Type IIB string/F-theory, it is known that 
the Gukov--Vafa--Witten superpotential \cite{GVW}, 
\begin{equation}
 W^{({\rm cpx})}(z) =  W_{GVW} = c \int_X G \wedge \Omega_X,
\label{eq:GVW}
\end{equation}
describes the effective superpotential of the complex structure moduli $z$ 
(including dilaton) for some coefficient $c$.
Non-perturbative effects may generate effective superpotential that 
depends on the K\"{a}hler moduli $T$; each one of such terms is of the form 
\begin{equation}
 W^{(T)}(T, z) = W_{np} = A(z) e^{- d T},  
\label{eq:gaugino-cond}
\end{equation}
and $d=2\pi/N$ in the case of gaugino condensation of 4D SU(N) super Yang--Mills theory; the prefactor $A$ may depend on the complex structure moduli 
field $z$ in principle, and at least it does on the choice of normalization 
of $\Omega_X$. Supersymmetric stabilization of the K\"{a}hler moduli is 
possible, if there are more than one term of 
the form (\ref{eq:gaugino-cond}) (the racetrack scenario), and the tuning (\ref{eq:cond-KL-min}) is assumed.
In order to work out the decompactification constraint on inflation discussed in the 
previous section, we need to know the coefficients $c$ and $A$.

The coefficient $c$ has been determined in \cite{Denef08} already. 
The procedure is to carry out dimensional reduction from 10D to 4D, and work out the 4D
scalar potential of complex structure moduli in flux compactification first. 
The value of $c$ is determined so that the 4D effective supergravity 
potential reproduces the potential obtained through the reduction. 
The result is that \cite{Denef08}
\begin{equation}
  c = M_{\rm Pl}^3 \frac{1}{\sqrt{4\pi}}
\label{eq:c-determined}
\end{equation}
for Type IIB orientifold on a Calabi-Yau 3-fold $X$,
when one uses the K\"{a}hler potential 
\begin{equation}
 \frac{K}{M_{\rm Pl}^2} =
   - \ln \left[ i \int_X \Omega \wedge \overline{\Omega} \right] 
     - \ln \left[ (\tau - \bar{\tau})/i  \right]
     - 2 \ln \left[ \frac{1}{3! g_s^{3/2}} \int_X \omega^3 \right].
\label{eq:eff-Kahler}
\end{equation}
Here, $\tau := C^{(0)} + i e^{- \phi} = C^{(0)} + i g_s^{-1} e^{-\tilde{\phi}}$ 
is the axion-dilaton chiral multiplet, $g_s$ the vacuum value of $e^\phi$, 
and $\omega$ the K\"{a}hler form\footnote{The K\"{a}hler form $\omega$ 
refers to the one in the Einstein frame metric $g_E$ of Type IIB 
10-dimensional supergravity; we assumed that the Weyl rescaling
from the string frame metric is given by 
$g_S \rightarrow e^{(\phi-\vev{\phi})/2} g_E$. We understand that $\omega $ has been made dimensionless by using the dimensionless coordinates $\underline{y}^m$.} 
on $X$. The last term is equivalent to the first term $-3 \ln({\rm Re}(T))$ 
in (\ref{eq:Kahler-noscale+cpx}), and the first two terms are identified 
with ${\cal K}^{({\rm cpx})}$.

We have adopted a convention to make everything dimensionless in 
$\int_X G \wedge \Omega$, by rendering the space coordinates $y^m$
dimensionless; $y^m \longrightarrow \underline{y}^m 
= y^m/\ell_s$, $\ell_s := 2\pi \sqrt{\alpha'} =: 1/M_s$. Three-form fluxes 
of Type IIB string theory are quantized as 
\begin{equation}
 \frac{1}{\ell_s^2} \int F^{(3)} \in \Z, \qquad 
 \frac{1}{\ell_s^2} \int H^{(3)} \in \Z,
 \label{flux}
\end{equation}
and are therefore turned into dimensionless integers 
$\int F^{(3)} = n^{R} \in \Z$ and $\int H^{(3)} = n^{NS} \in \Z$ 
in the dimensionless coordinate setting. Thus, 
\begin{equation}
 \int_X G \wedge \Omega =\sum_a (n^{R}_a - \tau n^{NS}_a) \Pi_a, 
\end{equation}
with integer flux quanta $n^R_a, n^{NS}_a \in \Z$ and dimensionless period 
integrals $\Pi_a$'s ($a=1, \cdots ,2(h^{2,1}_+ +1)$).

The prefactor $A$ in (\ref{eq:gaugino-cond}) for gaugino condensation 
is estimated by matching the moduli scalar potential from gaugino 
condensation in 4-dimensions to a {\it supergravity} potential of the 
effective theory of moduli fields. We work on the case of gaugino condensation 
of SU(N) super Yang--Mills theory originating from an SU(N) gauge group 
on a stack of N D7-branes. 

This is not the first time that this problem is addressed; 
Ref. \cite{temp2} did that in Heterotic string compactifications, for example. 
It will be easy to see that this problem involves subtlety, because 
there are non-decoupling effects in the $M_{\rm Pl} \longrightarrow \infty$ 
limit. Consider a factor $e^{K/M_{\rm Pl}^2}$, for example. It is just 1, if 
we simply set $M_{\rm Pl} \longrightarrow \infty$, but if we are to rewrite 
$K/M_{\rm Pl}^2$ by using (\ref{eq:eff-Kahler}), it remains a factor with 
non-trivial dependence on various moduli fields. When we use the results 
of gaugino condensation in the rigid supersymmetry limit, we need to be careful. 

To properly appreciate the subtleties in the context of Type IIB 
compactifications, let us start off by reminding ourselves of the well-known procedure of 
dimensional reduction. The Einstein--Hilbert term of the Type IIB 10D Einstein 
frame action is given by 
\begin{equation}
 S_{E} \supset  \frac{2\pi}{\ell_s^2 g_s^2} 
  \int d^4x \int_X d^6\underline{y} \sqrt{-g_E} R_E , 
\end{equation}
where the six coordinates on $X$ have been made dimensionless, and 
the Einstein frame metric $g_E$ is used. Dimensional reduction leads to 
\begin{equation}
 S_{E|4} \supset \frac{2\pi}{\ell_s^2 g_s^2} \int d^4 x \sqrt{-g_{E|4}} \;  
     \omega^3 R_{E|4} + \cdots, 
\end{equation}
where $g_{E|4}$ is the restriction of the 10D Einstein frame metric $g_E$. 
We call this action on 4-dimensions as that of {\it reduction frame}. 

This reduction frame action can be cast into an Einstein frame action 
by Weyl-rescaling. We adopt the following rescaling, 
\begin{equation}
  g_{E|4} \longrightarrow g_4 \times \frac{\vev{\omega}^3}{\omega^3}, 
\label{eq:rescale-red-E}
\end{equation}
so that all of the metrics $g_4$, $g_{E|4}$, $g_{E}$ and $g_S$ have the 
same normalization in their vacuum expectation values. 
With this Weyl rescaling, we obtain 
\begin{equation}
  S_{4} \supset \frac{4\pi \vev{\omega}^3}{2 \ell_s^2 g_s^2} 
   \int d^4 x \sqrt{-g_4} \; R_4, \qquad 
   M_{\rm Pl}^2 = \frac{4\pi \vev{\omega}^3}{ \ell_s^2 g_s^2 }. 
\end{equation}

It should be reminded that the reduction frame above is not the same as 
the frame that is used in writing down a 4D supergravity action on a 
superspace. Let a supergravity action on superspace be 
\begin{equation}
 S_{SS4} =  \int d^4x \int d^2 \Theta  2 {\cal E} \; \left[ 
   \frac{3M_{\rm Pl}^2}{8} \left(\overline{\mathscr{D}}^2 - 8R \right)
       e^{-\frac{K' + \Gamma}{3M_{\rm Pl}^2}}
   + W + H 2\tr {}_N[W^\alpha W_\alpha] \right] + {\rm h.c.},
\end{equation}
where $W$ is the superpotential, $K'$ the K\"{a}hler potential and  
$H$ the gauge kinetic function which may depend on chiral multiplets 
holomorphically; for all other notations, see \cite{Wess-Bagger}.
The Weyl rescaling from this superspace frame to the Einstein frame 
is 
\begin{equation}
 g_{SS4} \longrightarrow g_4 \times e^{\frac{K'}{3M_{\rm Pl}^2}}.
 \label{eq:rescale-SS-E}
\end{equation}
If we are to use the K\"{a}hler potential $K$ in (\ref{eq:eff-Kahler}) 
as $K'$ in the superspace, the Weyl rescaling factor 
in (\ref{eq:rescale-SS-E}) depends on the complex structure moduli fields, 
whereas the one in (\ref{eq:rescale-red-E}) does not. Even when it comes 
to the dependence on the K\"{a}hler moduli fields, the rescaling factor 
in (\ref{eq:rescale-SS-E}) is proportional to 
\begin{equation}
 e^{\frac{K}{3M_{\rm Pl}^2}} \propto e^{- \frac{2}{3}
    \ln \left[ \frac{\omega^3}{g_s^{3/2}} \right]} = \frac{g_s}{\omega^2},
\end{equation}
which is clearly different from the rescaling factor 
in (\ref{eq:rescale-red-E}). Note also that the vacuum values of the 
metric $g_{SS4}$ and $g_{E|4}$ in the two frames are not the same either, 
because $\vev{e^{K'/3M_{\rm Pl}^2}} \neq 1$, unless we choose 
$K' = K - {\rm const.}$, so that $\vev{K'} = 0$.

Reference \cite{KapLoui} discussed the matching of gaugino condensation between 
[4D Einstein / superspace frame action with a vector multiplet] and 
[4D effective theory of moduli fields (in the Einstein frame)], but  
the IIB-reduction frame was not used. We will thus translate the reduction 
frame action to the Einstein frame, and get the matching done, 
in the following.  

As is well-known, the reduction of DBI action of a stack of $N$ D7-branes 
gives rise to 
\begin{equation}
S_{E|4} \supset 
 \int d^4x d^2 \theta  \; \frac{T}{16\pi} 2{\rm tr}_N[W^\alpha W_\alpha]
   + {\rm h.c.}, \qquad 
  {\rm Re}(T) = \frac{\omega^2}{g_s} 
\label{eq:4D-gauge-kin-hol-super}
\end{equation}
in the IIB-reduction frame; the Weyl rescaling (\ref{eq:rescale-red-E}) 
keeps the gauge field as it is, but the gaugino also needs to be rescaled by 
\begin{equation}
\lambda_{E|4} \longrightarrow \lambda_{4,h} \times 
   \left(\frac{\omega^3}{\vev{\omega}^3} \right)^{\frac{3}{4}}
\end{equation}
so that the kinetic term is of the form 
\begin{equation}
  S_4 \supset \int d^4 x \sqrt{-g_4} \left[
  -\frac{1}{4g_{YM}^2} 2\tr[ F_{\mu\nu} F_{\rho \sigma} ] g^{\mu \rho}_4g^{\nu\sigma}_4
  -\frac{i}{g_{YM}^2} 2\tr[ \lambda_{4,h} \sigma^a e_a^{\; \mu} D_\mu \bar{\lambda}_{4,h}]
     \right]
\end{equation}
in the Einstein frame; the coefficient of $2{\rm tr}_N[W^\alpha W_\alpha]$ in $S_4$ is shifted to become 
\begin{equation}
 \frac{T}{16\pi} \longrightarrow \frac{1}{16\pi}
    \left[ T + \frac{2N}{2\pi}\frac{3}{4}
             \ln \left[\frac{\omega^3}{\vev{\omega}^3}\right]
    \right]
\end{equation}
because of the rescaling anomaly; $2N$ is the number of gaugino zero modes 
in a 1-instanton background. The relation 
\begin{equation}
   \langle {\rm Re}(T) \rangle = \frac{4\pi}{g^2_{\rm YM}}
   \label{eq:T-gYM-relation}
\end{equation}
remains the same, however. 

The well-known result of gaugino condensation in rigid supersymmetry, 
\begin{equation}
    \frac{\langle 2 {\rm tr}_N[\lambda \lambda]_h \rangle}{32\pi^2} =
    N \Lambda_h^3 ,
\label{eq:gaugino-cond-vev}
\end{equation}
should be understood as follows.  
First, gaugino $\lambda_{4,h}$ in the Einstein frame action with holomorphic 
normalization is used on the left-hand side, since it is the Einstein frame 
metric $g_4$ whose vacuum value becomes $\eta_{\mu \nu}$ (rather than 
$g_{SS4}$ in the superspace frame), and that such terms as 
$[K_T (\partial_\mu T)]$ in the covariant derivative of gaugino 
$D_\mu \lambda$ do not play a role in determining the instanton zero mode 
configuration \cite{temp2, KapLoui}. 

Secondly, the dynamical scale $\Lambda_h$ is given by 
$\Lambda_h^{3N} = \mu^{3N} e^{-\frac{8\pi^2}{g_{YM}^2(\mu)} + i \theta}$ using the 
gauge coupling $g_{YM}^2(\mu)$ in the Einstein frame action 
renormalized at scale $\mu$. In terms of string compactification, 
\begin{equation}
   \Lambda_h^3 \simeq M_{KK}^3 e^{-\frac{2\pi}{N}T}.
\label{eq:Lambdah-T-relation}
\end{equation}
Some power of $[{\rm Re}(T)/{\rm Re}\vev{T}]$ may be included on the right-hand side, but it is not 
important for the purpose of determining the prefactor $A$ through matching.
The renormalization scale $\mu$ was replaced by the Kaluza--Klein scale 
$M_{KK}$, because the D7-brane DBI action on a flat spacetime background 
is a good approximation only at energy scale above 
$M_{KK}$ \cite{temp2}.\footnote{\label{fn:threshold}
When 1-loop threshold corrections are included in the 
relation (\ref{eq:T-gYM-relation}), the factor $M_{KK}^3$ is replaced 
by a more precise expression in the form of 
$[1/\ell_s \sqrt{\vev{\omega}}]^3$ times some dimensionless ``value'' 
that is generically of order unity. 
Since spectrum around the Kaluza-Klein scale depends on complex structure moduli $z$, so does the 1-loop threshold correction. 
The ``value'' should therefore actually be a function  of $z$; this is how the $z$-dependence originates in 
the gaugino condensation superpotential $W^{(T)}(T,z)$.}

The Einstein frame 4D supergravity action in a theory with an $\SU(N)$ vector 
multiplet has such terms as \cite{Wess-Bagger}
\begin{eqnarray}
  \sqrt{-g_4}^{-1} {\cal L}_4 & \supset  & e^{K/M_{\rm Pl}^2} \left[
    3 \left| \frac{W}{M_{\rm Pl}} \right|^2 \right.  \label{eq:vector-sugra} \\
  & & \qquad \left. 
 - K^{\bar{j}i} 
     \left[ D_iW - H_{,i} e^{-\frac{K}{2M_{\rm Pl}^2}} 2\tr[\lambda \lambda]_{4,h}\right]
     \left[ \overline{D_jW} - \overline{H_{,j}} e^{- \frac{K}{2M_{\rm Pl}^2}}
              2 \tr [\bar{\lambda}\bar{\lambda} ]_{4,h} \right]
    \right],  \nonumber 
\end{eqnarray}
and the gauge kinetic function $H$ is given by $T/(16\pi)$ in our context; 
the gauge kinetic function remains the same under the Weyl rescaling 
(\ref{eq:rescale-SS-E}) between the superspace frame and Einstein frame.
The idea of \cite{KapLoui} for the matching is to replace the gaugino 
composite operators in (\ref{eq:vector-sugra}) by the moduli dependent 
expectation value (\ref{eq:Lambdah-T-relation}), and come up with 
an effective superpotential $W_{\rm eff.}(T)$ so that the auxiliary F-term 
of the K\"{a}hler moduli chiral multiplet $T$ reproduces the potential 
generated from (\ref{eq:vector-sugra}). 

We find that 
\begin{equation}
 W_{\rm eff.} \sim W + M_{KK}^3 N^2 e^{-\frac{\vev{K}}{2M_{\rm Pl}^2}} \; e^{-\frac{2\pi}{N}T}.  
\label{eq:matching-1}
\end{equation}
To see this, it would not be difficult to see that a simple computation 
\begin{equation}
  - (\partial_T H) 2\tr {}_N [\lambda \lambda]_{4,h} e^{-\frac{K}{2M_{\rm Pl}^2}}
 = - 2\pi N M_{KK}^2 e^{-\frac{2\pi}{N}T} e^{-\frac{K}{2M_{\rm Pl}^2}}
 = \partial_T \left[ N^2 M_{KK}^3 e^{-\frac{2\pi}{N}T} \right] e^{-\frac{K}{2M_{\rm Pl}^2}}
\label{gauginomatch}
\end{equation}
is behind the identification of $W_{\rm eff.}$, first of all. Secondly, 
we should remember that this matching is ultimately based on the the result 
of rigid supersymmetry (\ref{eq:gaugino-cond-vev}), and we can determine how 
$\Lambda_h^3$ depends on $T$ and $\vev{T}$ separately only through guess work.
We chose to deal with $M_{KK}$ and $e^{-\vev{K}/2M_{\rm Pl}^2}$ 
in (\ref{eq:matching-1}) as their vacuum values at the end, not as 
field-dependent functions, because we know that the effective superpotential 
of gaugino condensation depends on the K\"{a}hler moduli {\it field} $T$ 
only in the form of $e^{-\frac{2\pi}{N}T}$.

Note that the new term from gaugino condensation in the effective theory 
superpotential $W_{\rm eff.}$ in (\ref{eq:matching-1}) also contributes to 
the $-3|W_{\rm eff.}/M_{\rm Pl}|^2$ term in the effective theory scalar potential. 
This contribution in the effective theory cancels against the 
$K^{\bar{T}T}|K_T W_{\rm eff.}|^2$ term in the effective theory scalar potential, 
which we kept out of (\ref{gauginomatch}), provided the 
K\"{a}hler potential of $T$ is in the no-scale type. Cancellation also takes 
place on the other side of the matching: there is no such term as 
\begin{equation}
\frac{|H|^2}{M_{\rm Pl}^2}
e^{-K/3M_{\rm Pl}^2} \tr[\lambda \lambda]_{4,h} \tr[\bar{\lambda}\bar{\lambda}]_{4,h}
\end{equation}
in the Einstein frame action $\sqrt{-g_4}^{-1} {\cal L}_4$ of supergravity with 
an $\SU(N)$ vector multiplet \cite{Wess-Bagger}. Thus, there is no 
contradiction in the current set-up, when we adopt the matching condition 
(\ref{eq:matching-1}).

The effective superpotential of the gaugino condensation is rewritten into 
a form that fits better for practical analysis.
Noting that 
\begin{equation}
 M_{KK} \sim \frac{1}{\ell_s \sqrt{\vev{\omega}}} \sim 
  M_{\rm Pl} \frac{1}{\sqrt{4\pi}} \frac{g_s}{\vev{\omega}^2}
  = M_{\rm Pl} \frac{1}{\sqrt{4\pi} {\rm Re}\vev{T}},
\end{equation}
and that 
\begin{equation}
 e^{-\frac{\vev{K}}{2M_{\rm Pl}^2}} \sim \left[ \int \Omega \; \sqrt{2/g_s} \right] 
   \times \left[{\rm Re}\vev{T}\right]^{3/2},
\end{equation}
we rewrite the gaugino condensation contribution to $W_{\rm eff.}$ as 
\begin{equation}
W_{np}(T,z) = M_{\rm Pl}^3 \frac{a \; N^2}{[4\pi {\rm Re}\vev{T}]^{3/2} }
      e^{-\frac{2\pi}{N}T}, 
\end{equation}
with 
\begin{equation}
  a \sim  \left[ \int \Omega \times \sqrt{2/g_s} \right].
\end{equation}
The 1-loop threshold correction to the relation (\ref{eq:T-gYM-relation}), 
which generically depends on the complex structure moduli fields $z$, is 
implemented in this factor $a$.  

We now therefore conclude that the superpotential of the moduli effective 
theory is of the form 
\begin{equation}
 W_{\rm eff} = M_{\rm Pl}^3 \left[
    \sum_i \frac{1}{\sqrt{(4\pi \vev{{\rm Re}(T)})^3}} a_i N_i^2 e^{ - \frac{2\pi}{N_i} T}
   + \frac{1}{\sqrt{4\pi}} \sum_a (n^R_a-\tau n^{NS}_a) \Pi_a \right],
\end{equation}
where $a_i$'s are dimensionless and will remain of order unity, though 
it may depend on the complex structure moduli fields $z$'s, and on 
normalization of $\Omega_X$.

As a sanity check, we use the effective superpotential (\ref{eq:GVW}, 
\ref{eq:c-determined}) and the effective K\"{a}hler potential 
(\ref{eq:eff-Kahler}) to estimate physical masses of the complex structure 
moduli, and see whether the result is sensible. 
Ignoring the complex structure moduli dependence of the coefficient $A$ 
in (\ref{eq:gaugino-cond}), the physical mass of complex structure moduli 
can be estimated by using 
$m^2_{z\bar{z}} \sim e^{K/M_{\rm Pl}^2} |K^{\bar{z}z} \partial_z^2 W|^2$ 
[cf. (\ref{eq:phys-mass})]:
\begin{eqnarray}
  m^2_{z\bar{z}} & \sim & \left[ 
  \frac{1}{ i \int_X \Omega_X \wedge \overline{\Omega}_X}
    \frac{g_s}{2}\frac{g_s^3}{\vev{\omega}^6} \right]
    \left[ \frac{\int_X \chi _a \wedge \bar{\chi}_b }
                {\int_X \Omega \wedge \overline{\Omega}} \right]^{-2}
    \frac{1}{4\pi} 
    \left| \sum_a (n^R_a-\tau n^{NS}_a) \partial_z^2 \Pi_a \right|^2 M_{\rm Pl}^2,
     \nonumber \\
  & \sim & \frac{M^2_{\rm Pl}}{4\pi} \frac{g_s^{3}}{\vev{\omega}^6} 
   \sim \frac{g_s}{\ell_s^2 \vev{\omega}^3} 
   \sim \left[ \frac{g_s^{1/2} M_{KK}^3}{(1/\ell_s)^2} \right]^2,
\end{eqnarray}
where $\{ \chi _a \text{'s} \}$ is a basis of (2,1) forms on a Calabi-Yau 3-fold $X$. When passing to the second line, we evaluated $(n^R_a-\tau n^{NS}_a)$ as $g_s^{-1/2}$ (somewhere in between 
$1$ and $g_s^{-1}$), and $|\partial_z^2 \Pi|^2/
[\int \Omega_X \wedge \overline{\Omega}_X]$ as $\mathcal{O}(1)$. 
This reproduces the mass estimate $(M_{KK}^3/M_s^2)$ of the complex structure 
moduli in \cite{KST}, passing the first sanity check.\footnote{The physical 
mass $m^2_{z\bar{z}}$ comes out to be hierarchically smaller than $M_{\rm Pl}^2$  
because of $e^{\vev{K}/M_{\rm Pl}^2} \ll 1$ in the convention in this article. 
Due to the K\"{a}hler--super Weyl transformation in 4D ${\cal N}=1$ 
supergravity (e.g., arbitrariness of the normalization of $\Omega_X$), 
both the superpotential and the K\"{a}hler potential need to be 
included in order to obtain physical $m^2_{z\bar{z}}$.
} 

The physical mass of K\"{a}hler moduli $m_{T\bar{T}}^2$, on the other hand, 
is estimated by 
\begin{eqnarray}
  m_{T\bar{T}}^2 & \sim & 
    \left[ \frac{1}{i \int_X \Omega \wedge \overline{\Omega}} 
           \frac{g_s}{2} \frac{1}{{\rm Re}(\vev{T})^3} \right]
    \left[ \frac{{\rm Re}(\vev{T})^2}{3} \right]^2 
    \frac{ 
      \left| \sum_{i=1,2} a_i(z) (N_id_i)^2 e^{-d_i \vev{T}} \right|^2 }
       {[(4\pi){\rm Re}(\vev{T})]^3 } M_{\rm Pl}^2
    \nonumber \\
  & \sim & \frac{(2\pi)^4 M_{\rm Pl}^2}{(4\pi)^3 {\rm Re}(\vev{T})^2 }
    \sim \left[\frac{(2\pi)^2}{4\pi \ell_s \sqrt{\vev{\omega}}} \right]^2
    \sim \left[ \pi M_{KK} \right]^2;
\end{eqnarray}
we ignored the complex structure moduli dependence of $a_i$'s in 
(\ref{eq:phys-mass}), so that the mixing terms disappear, and 
$m_{T\bar{T}}^2 \sim e^{K/M_{\rm Pl}^2} | K^{\bar{T}T} \partial^2_T W|^2$ is used for 
the estimate. 
This result is acceptable, in that the physical mass should not be higher 
than the Kaluza--Klein scale (where 4D ${\cal N}=1$ super Yang--Mills is 
not a good approximation). Thus, we do not find anything counter intuitive 
in the result of the physical mass $m^2_{T\bar{T}}$ of the K\"{a}hler moduli.
In reality, the value of $| \sum_i a_i e^{-d_i \vev{T}} |^2/ 
[(i \int_X \Omega \wedge \overline{\Omega}) \cdot (2/g_s)]$---a factor 
treated as ${\cal O}(1)$ in the second line---will be smaller than unity 
because of the exponential factors (see the next section for more). 
Depending on the exponential factors, diagonal entries of the physical 
mass matrix may be larger in $m^2_{T\bar{T}}$ or in $m^2_{z\bar{z}}$;  
they can also be comparable in size. 

The discussion so far indicates that $|W^{(T)}| \ll |W^{({\rm cpx})}|$ generically 
(in the moduli space) in Type IIB/F-theory compactifications, unless a 
dedicated condition like (\ref{eq:cond-KL-min}) is imposed for phenomenology.  
However, $m_{T\bar{T}}^2$ can be larger than $m_{z\bar{z}}^2$, in the case 
exponential factors are not too small, because of 
$M_{\rm Pl}^2 \vev{K^{\bar{T}T}} \sim ({\rm Re}\vev{T})^2 \gg 1$. 
In the large complex structure region of the moduli space, though, 
the physical mass $m_{z\bar{z}}^2$ may also be enhanced by 
$M_{\rm Pl}^2 K^{\bar{z}z} \gg 1$.

1-loop threshold corrections to the gauge coupling have been computed 
explicitly for some set-ups of string compactifications. Based on 
such known forms of the threshold corrections, it is reasonable to 
assume that $\vev{\partial_z a(z)}$ is not much different from $\vev{a(z)}$, 
both being dimensionless values of order unity. For this reason, 
$\vev{\partial_T \partial_z W}$ in the off-diagonal entries of the 
mass matrix (\ref{eq:phys-mass}) is not expected to be much smaller than 
$\vev{\partial_T^2 W}$ in the diagonal entry. We have used this expectation 
already in the discussion on reheating process at the end of 
the previous section. 

\section{Robust Estimate of Decompactification Constraint}
\label{sec:barrier}

Having studied the prefactor of the effective superpotential 
from gaugino condensation, let us now study how the K\"{a}hler moduli 
stabilization requirement (which we discussed in section \ref{sec:KL-problem}) 
constrains the possibility of inflationary process involving complex 
structure moduli. In this section and in the rest of this article, we stick 
to systems with effectively only one K\"{a}hler modulus field for simplicity,   
with the no-scale type K\"{a}hler potential (see footnote~\ref{fn:multi-h11}), 
and use the racetrack superpotential from two gaugino condensations. 

Racetrack superpotential has long been studied in the literature 
for volume stabilization \cite{Krasnikov, Escoda:2003fa}. 
In our context, where gaugino condensation of $\SU(N_1) \times \SU(N_2)$ 
contributes to the racetrack superpotential,
\begin{align}
 W^{(T)} &=  \frac{M_{\rm Pl}^3}{[(4\pi) {\rm Re}\vev{T}]^{3/2}}
   \left( a_1 N_1^2 e^{-\frac{2\pi}{N_1} T} + a_2 N_2^2 e^{-\frac{2\pi}{N_2} T} \right),
\label{racetracksuperpotential}
\end{align}
where $a_1, a_2$ are dimensionless and may depend on the complex structure 
moduli.

\subsection{Barrier Height of Volume Stabilization Potential}
\label{ssec:racetrack}
We begin with a review of important properties of this superpotential. 
In this section \ref{ssec:racetrack}, 
we take a close look at an
effective theory where all the complex structure moduli fields 
are replaced by their vacuum values. 
This is to take 
\begin{align}
W^{(\text{tot})}=W^{(T)}(T)+W_0,
\label{Wtot}
\end{align}
and the Kallosh--Linde tuning implies that 
\begin{align}
W_0 =-\vev{W^{(T)}}.
\label{eq:W0-def}
\end{align}

From a practical perspective, it is best to study how the scalar potential 
for the K\"{a}hler modulus $T$ changes, for different vacuum values $\vev{T}$. 
With a simplifying ansatz that 
\begin{equation}
  N_1 = N_2 + 1,
\label{eq:N1=N2+1}
\end{equation}
the vacuum value $\vev{T}$ is determined by 
\begin{align}
\vev{T} = \frac{N_1(N_1-1)}{2\pi}
      \ln \left[-r \frac{N_1-1}{N_1} \right] 
\label{eq:T-r-N-relation}
\end{align}
as a function of $N_1$ and a ratio $r := \vev{a_2/a_1}$, 
and then $W_0$ is given by
\begin{align}
W_0 =-\vev{W^{(T)}}=
-\frac{M_{\text{Pl}}^3}{[(4\pi )\Re \vev{T}]^{3/2}}
\vev{a_1}N_1
 \left(-\frac{\vev{a_1}}{\vev{a_2}}\frac{N_1}{N_2}\right)^{N_2} .  
\end{align}
\begin{figure}[tbp]
 \begin{center}
  \begin{tabular}{c}
  \includegraphics[width=.4\linewidth]{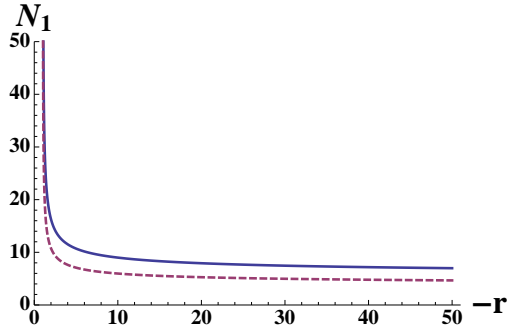}  
  \end{tabular}
  \caption{\label{fig:r-N-barrier-1} 
This is a contour plot of the vacuum value $\vev{T}$ on the 
$(-r, N_1)$ plane; the curve above (solid) is the contour of 
$\vev{T} = 25$, and the one below (dashed) that of $\vev{T}=10$.  }
 \end{center}
\end{figure}
Figure~\ref{fig:r-N-barrier-1} shows how the required value of 
$N_1$ changes for different values of $r = \vev{a_2/a_1}$ in order to 
achieve a given value of ${\rm Re}\vev{T}$. 

In supersymmetric extensions of the Standard Model with supersymmetry 
breaking around the TeV--a few hundred TeV, the unified gauge coupling 
constant is approximately $1/\alpha_{\rm GUT} \sim 25$; if there are 
$\SU(5)_{\rm GUT}$-charged particles in the light spectrum (e.g., in 
gauge mediated supersymmetry breaking scenario), however, the unified 
coupling constant may be stronger. It is not crazy to think also of 
$1/\alpha_{\rm GUT} \sim 10$ for this reason. When the gauge groups of the 
supersymmetric Standard Models originate from 7-branes wrapped on a 4-cycle 
in a Calabi--Yau 3-fold, such values of $1/\alpha_{\rm GUT} = 4\pi/g_{\rm GUT}^2$
can be used as the vacuum value of ${\rm Re}\vev{T}$. The two contours 
in Figure~\ref{fig:r-N-barrier-1} are drawn for $\vev{T} = 25$ 
and $\vev{T} = 10$ for this reason.\footnote{
\label{fn:fit-detail}
It is possible that there are a couple of integer factors entering 
various formulas we have used so far. First, depending on the intersection 
ring of a Calabi--Yau 3-fold for Type IIB compactification, 
$M_{\rm Pl}^2 \ell_s^2 = (4\pi) M_0 \vev{\omega}^3/g_s^2$ with $M_0 \ge 1$.
Secondly, depending on which topological cycle the GUT 7-branes are wrapped on, 
$1/\alpha_{\rm GUT} = M_1 \vev{\omega}^2/g_s$. This factor also affects 
the Kaluza--Klein scale on the GUT 7-brane: $M_{KK; GUT} = 
1/[\ell_s \sqrt{\vev{\omega}} M_2]$. Combining them together, 
\begin{equation}
\left(\frac{M_{\rm Pl}}{M_{KK;GUT}}\right)^2 = M_0 M_2^2 (4\pi)[{\rm Re}\vev{T}]^2
 = M_0 (M_2/M_1)^2 \frac{4\pi}{[\alpha_{\rm GUT}]^2}.
\end{equation}
Because of the fudge factor $M_0 (M_2/M_1)^2$, it does not immediately run 
into inconsistency to choose $M_{KK; GUT} \simeq ({\rm a~few}) \;
10^{16} \GEV$, while taking $1/\alpha_{\rm GUT} \sim 10$, for example. 
One should keep in mind, however, that there is such a tension. 
See also footnote~\ref{footnote3}. For the parameter fitting in 
the case of F-theory compactifications, see \cite{RHnu-Ftheory}.
}  

It is worth noting that, for any choice of the parameters $\vev{a_2/a_1}$, 
$N_{1,2}$ of the effective theory, the potential barrier height of the 
K\"{a}hler modulus in 
\begin{equation}
 V_{\rm racetrack}(T) := \frac{1}{|a_1|^2} \; e^{-3 \ln[ {\rm Re}(T) ] } 
  \left[ K^{\bar{T}T} |D_T W^{(\text{tot})}|^2
        -3 \left|\frac{W^{(\text{tot})}}{M_{\rm Pl}}\right|^2 \right] 
\label{eq:Vracetrack-def}
\end{equation}
using $W^{(\text{tot})}$ in (\ref{Wtot}) is bounded from above by 
\begin{eqnarray}
 [V_{\rm racetrack}]^{\rm barrier} & \approx & 
  \frac{1}{ [{\rm Re}\vev{T}]^3 }
  \frac{[{\rm Re}\vev{T}]^2}{ M_{\rm Pl}^2 }
  \frac{M_{\rm Pl}^6}{(4\pi)^3 [{\rm Re}\vev{T}]^3} \times 
   \left| (2\pi/N_i)N_i^2 e^{- \frac{2\pi}{N_i}\vev{T}}\mbox{'s} \right|^2,  \nonumber \\
  & \approx & \frac{M_{\rm Pl}^4}{(4\pi)^3 [{\rm Re}\vev{T}]^4}
   \left| (2\pi N_{i}) e^{- \frac{2\pi}{N_i} \vev{T}}\mbox{'s} \right|^2 
  \approx \frac{M_{KK}^4}{4\pi} 
      \left|(2\pi N_{i}) e^{- \frac{2\pi}{N_i} \vev{T}}\mbox{'s} \right|^2. 
 \label{eq:racetrack-barrier-height-estimate}
\end{eqnarray}
It cannot exceed the energy scale $(M_{KK})^4$ by much, obviously from 
construction, and is further suppressed by exponential factors. 

\begin{figure}[tbp]
 \begin{center}
  \begin{tabular}{cc}
  \includegraphics[width=.5\linewidth]{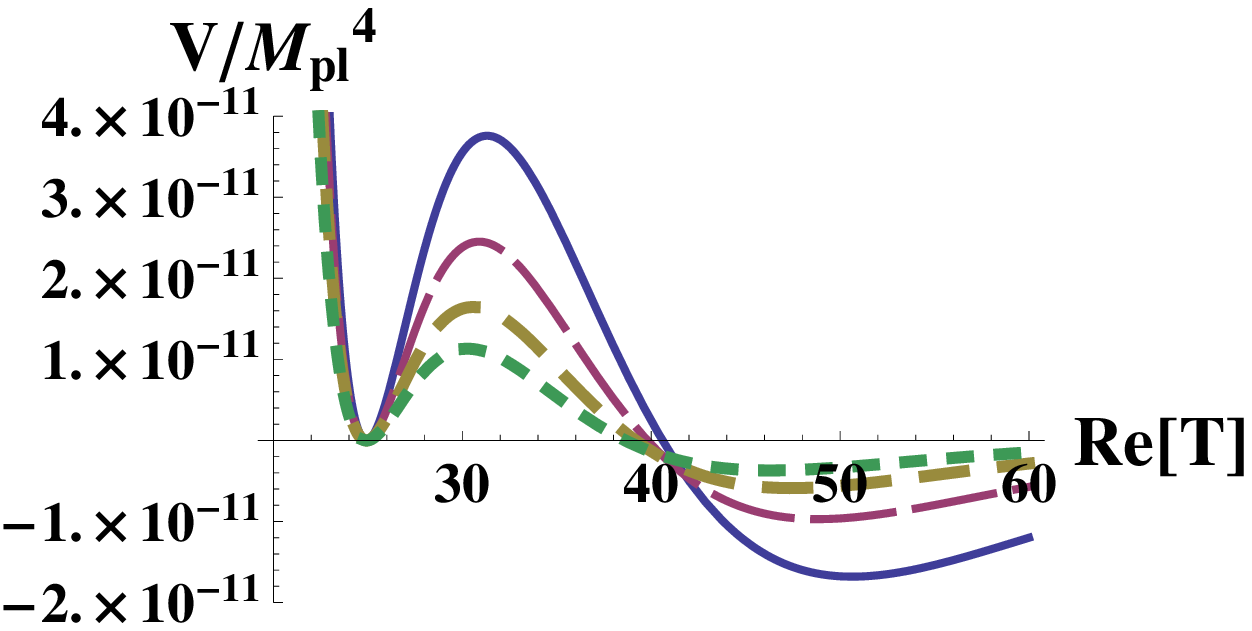} &
  \includegraphics[width=.5\linewidth]{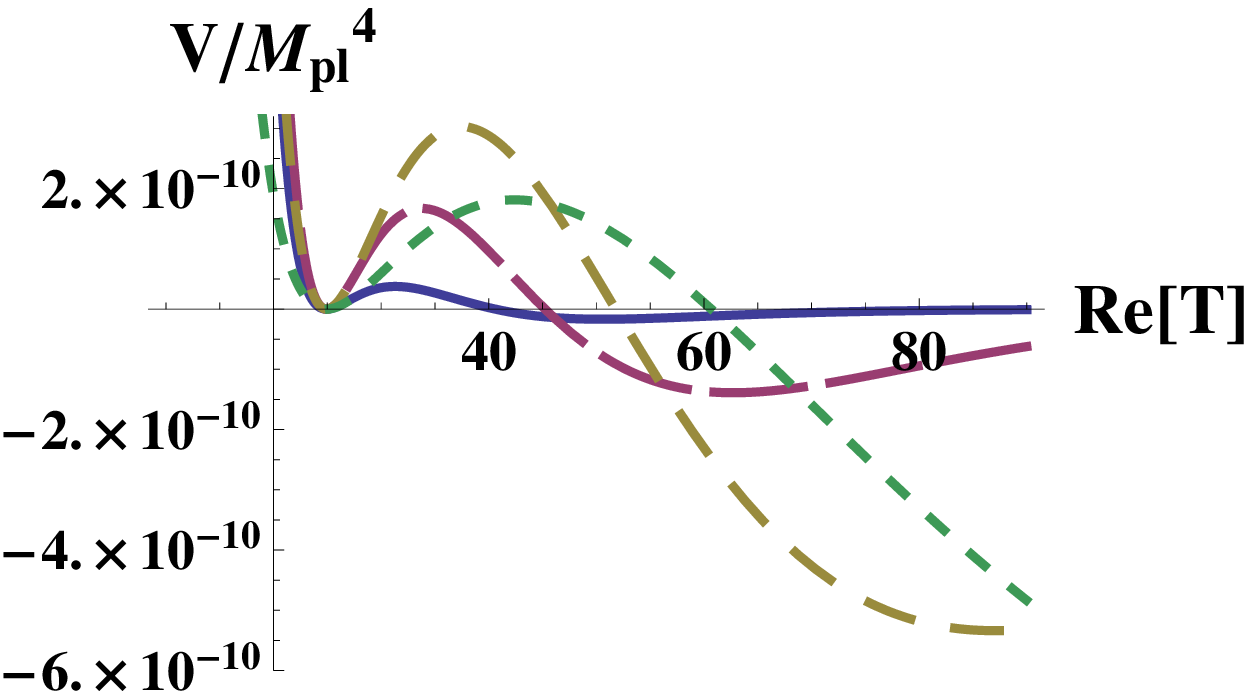} \\
  (a) & (b) 
  \end{tabular}
  \caption{
\label{fig:r-N-barrier-2} 
The potential $V_{\rm racetrack}(T)$ for a couple 
of different choices of $r := \vev{a_2/a_1}$ and $N_1$ that lead to 
$\vev{T} = 25$.
{\bf Panel (a):} the curves shown in solid, long dashed, dashed and dotted 
lines are for parameters $(r,N_1)=(-1.05,68)$, $(-1.06,62)$, $(-1.07,57)$ and 
$(-1.08,53)$, respectively. The larger the value of $r$ is, the larger  
the height of the potential barrier becomes.
{\bf Panel (b):} the potential for the values of $(r,N_1) = (-1.05,68)$, 
$(-1.02,118)$, $(-1.007,238)$ and $(-1.002,626)$ is shown in 
a solid, long dashed, dashed and dotted line, respectively.
The second (negative energy) minimum deepens for larger $r$. 
}
 \end{center}
\end{figure}

Figure~\ref{fig:r-N-barrier-2} shows the potential $V_{\rm racetrack}(T)$ 
for a couple of different choices of $(r, N_1)$ satisfying 
${\rm Re}\vev{T} = 25$. The shape of the volume stabilizing potential 
$V_{\rm racetrack}(T)$ and the height of the barrier are very sensitive to 
the choice of $(r, N_1)$.
This high sensitivity of the barrier height is also an unavoidable 
consequence in the racetrack potential. To see this, note that 
the height of the potential barrier, $[V_{\rm racetrack}]^{\rm barrier}$, is 
qualitatively explained by the two factors: one is 
\begin{equation}
  \frac{(M_{KK})^4}{4\pi} \simeq 
  \frac{M_{\rm Pl}^4}{(4\pi)^3 ({\rm Re}\vev{T})^4} \simeq 
  1.3 \times 10^{-9} \left(\frac{25}{{\rm Re}\vev{T}}\right)^4 M_{\rm Pl}^4,
\label{eq:barrier-factor-1}
\end{equation}
and the other is from the extra exponential factors 
in (\ref{eq:racetrack-barrier-height-estimate}); the latter is evaluated to be\footnote{In reality, the two 
factors---(\ref{eq:barrier-factor-1}) and (\ref{eq:barrier-factor-2}) 
combined---almost explains the hierarchy 
$[V_{\rm racetrack}]^{\rm barrier} \ll M_{\rm Pl}^4$ obtained numerically, 
but unaccounted hierarchy still remains. For example, 
for the choice $(r, N_1) \simeq (-1.05, 68)$, we found 
$[V_{\rm racetrack}]^{\rm barrier} \sim 10^{-11} \times M_{\rm Pl}^4$ 
numerically [Figure~2(a)], whereas the two factors (\ref{eq:barrier-factor-1},
\ref{eq:barrier-factor-2}) combined predicts $10^{-9} \times 10^{3}$.
Similarly for $(r, N_1) \simeq (-1.67, 20)$, the numerical results was 
$[V_{\rm racetrack}]^{\rm barrier}/M_{\rm Pl}^4 \simeq 10^{-15}$ [Figure~3] ,
which is still smaller than the estimate $10^{-9} \times 10^{-3}$. 
The combination of (\ref{eq:barrier-factor-1}) and (\ref{eq:barrier-factor-2})
tends to overestimate $[V_{\rm racetrack}]^{\rm barrier}/M_{\rm Pl}^4$ presumably 
because there is cancellation between the two exponential factors, and 
also because the barrier height should have been estimated by using a value of 
$T$ somewhat larger than $\vev{T}$.
}  
\begin{equation}
 \left| (2\pi N_1)e^{- \frac{2\pi}{N_1} \vev{T}} \right|^2_{\vev{T}=25} \simeq 
   \left\{ 
   \begin{array}{ll}
      1800 & (r, N_1) \simeq (-1.05, 68), \\
      0.002^{} & (r, N_1) \simeq (-1.67, 20). 
   \end{array}
   \right.
\label{eq:barrier-factor-2}
\end{equation}
For the value of $r = - \vev{a_2/a_1}$ close to $-1$, where 
$\ln[-r] \simeq 0$, the value of $N_1$ for a given $\vev{T}$ changes rapidly, 
as seen in (\ref{eq:T-r-N-relation}) or in Figure~\ref{fig:r-N-barrier-1}, and 
hence the exponential factor becomes small very quickly as the value of 
$r$ differs from $-1$. For larger value of $|r|$, $N_1$ does not change much, 
so that the rapid decrease in the barrier height slows down 
(Figure~3). To summarize, the potential barrier 
$[V_{\rm racetrack}]^{\rm barrier}/M_{\rm Pl}^4$ is highly sensitive to the 
choice of the parameter $r = - \vev{a_2/a_1}$, because the value of 
$\vev{T}/N_1$ depends very much on $r$, and further because it is 
exponentiated. 
\begin{figure}[tbp]
 \begin{center}
  \begin{tabular}{c}
  \includegraphics[width=.5\linewidth]{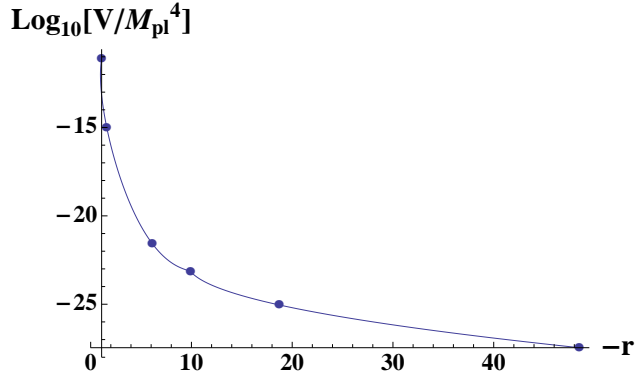}  
  \end{tabular}
  \caption{The potential barrier height 
$[V_{\rm racetrack}]^{\rm barrier}/M_{\rm Pl}^4$ changes by orders of magnitude 
for different choices of $(r, N_1)$. The data points are for 
$(r, N_1) \simeq (-1.1, 50)$, $(-1.67, 20)$, $(-6.1, 10)$, $(-9.9, 9)$, 
$(-18.7,8)$, and $(-48.5, 7)$.}
 \end{center}
\end{figure}

The potential barrier height can remain close to $M_{KK}^4$, when the value 
of $\ln[-r] = \ln[-\vev{a_2/a_1}]$ is very small, obviously from the 
discussion above. Even in an extreme choice\footnote{For $r$ even closer to 
$-1$, however, the potential barrier height is not as high as in the case 
with $r=-1.007$; see Figure~\ref{fig:r-N-barrier-2}~(b). We have not looked 
into the details of what is going on.} $(r, N_1) \simeq (-1.007, 238)$, 
however, the barrier height remains to be roughly 9.5 orders of magnitude 
below $M_{\rm Pl}^4$, or equivalently, 
$[V_{\rm racetrack}]^{\rm barrier} \sim (10^{16} \; \GEV)^4$. 
When the parameter $r$ is not as finely tuned as above, the energy scale of 
the barrier height, $([V_{\rm racetrack}]^{\rm barrier})^{1/4}$, is already five or 
six orders of magnitude below the Planck scale $M_{\rm Pl}$. 

We consider that it is a generic consequence of the racetrack superpotential 
from 4D gaugino condensations in the regime $\vev{T} \approx {\cal O}(10)$ 
or somewhat larger, 
even with the Kallosh--Linde tuning (\ref{eq:cond-KL-min}), that 
the height of the volume stabilizing barrier is no higher than $M_{KK}^4$, 
and it generically comes with extra exponential suppression. The exponential 
suppression is mitigated only with very dedicated choice of parameters 
$(\vev{a_2/a_1}, N_1)$. We have also done the same analysis for $\vev{T} = 10$ 
as well, and found similar results; $[V_{\rm racetrack}]^{\rm barrier}$ can only 
be as high as $({\rm a~few} \times 10^{16} \; \GEV)^4$ for an extreme 
choice of $(\vev{a_2/a_1}, N_1)$. It is likely that the same argument holds 
true for any other mechanism for $W^{(T)}$, as long as it relies on 
non-perturbative dynamics in the 4D effective gauge theory below the 
Kaluza--Klein scale. 

Certainly one cannot use the barrier height of the potential 
$V_{\rm racetrack}(T)$ alone to exclude some kinds of inflation models, since 
the potential $V_{\rm racetrack}(T)$ in (\ref{eq:Vracetrack-def}) is not 
the same as the scalar potential of the full 
theory (\ref{eq:scalar-potential-ZandT}). It has been accepted as a rule 
of thumb, however, that the volume stabilization may be in jeopardy, 
if the vacuum energy during inflation is comparable to the potential barrier 
height of $V_{\rm racetrack}(T)$. 

Therefore we have an important lesson, under this ``rule''.
In the regime of moderately large volume, i.e., ${\rm Re}\vev{T} \sim 10$ 
or somewhat larger, it is very hard to accommodate inflation with the 
energy density $\rho \sim (10^{16} \; \GEV)^4$ or larger; this energy density 
corresponds to the tensor-to-scalar ratio that can be probed in a near future 
($10^{-1}\mbox{--}10^{-2}$ or so). Depending on the value of the 
tensor-to-scalar ratio, one might be motivated to think of extra light matter 
particles (so that the unified gauge coupling is stronger) and/or Kaluza--Klein 
scale higher than the energy scale of apparent gauge coupling unification, 
or to throw away the scheme of supersymmetric unification altogether.
Note that this statement is very robust,\footnote{
It should be reminded, though, that our numerical study is only for 
a pair of gaugino condensations, and a relation (\ref{eq:N1=N2+1}) is imposed 
just to keep the presentation simple. Details in footnote \ref{fn:fit-detail}
should also be taken into account.} because it only refers to the energy 
density during inflation, not to details of inflation models. 

In the analysis of section \ref{ssec:phen-limit-destabil}, we use the 
following parameter set (like in \cite{KL-problem}):
\begin{align}
N_1=68,\quad N_2=67,\quad \vev{a_2/a_1}\simeq -1.05041,
\label{eq:vac-parameter}
\end{align}
when $\vev{T}=25$ and $[V_{\rm racetrack}]^{\rm barrier} \simeq 4 \times 
10^{-11} \times M_{\rm Pl}^4$ (the tuning (\ref{eq:cond-KL-min}) is understood). 
This choice of parameters is less extreme than 
$(\vev{a_2/a_1}, N_1)=(-1.007,238)$, so we expect to derive a more robust 
constraint on the complex structure deformation allowed during the 
inflationary process. The other (and more important) reason is that 
the potential $V_{\rm racetrack}(T)$ develops a deep negative energy minium in the region 
${\rm Re}(T) \gg \vev{T} = 25$, if $\vev{a_2/a_1}$ is chosen to be 
close to the value $-1.007$ for the highest $[V_{\rm racetrack}]^{\rm barrier}$ 
possible (see Figure~2(b)). We need to be worried about the quantum tunneling in that case. The choice above is more moderate 
from that perspective. Although this quantum stability argument should also 
be included in the discussion on the tensor-to-scalar ratio above, we do not 
try to conduct quantitative analysis in this article.

\subsection{Phenomenological Study of Volume Destabilization using 
$V_{\rm racetrack}$}
\label{ssec:phen-limit-destabil}

If we are to seek for a case where the complex structure moduli sector 
$z$ plays a non-negligible role during inflation, then 
$W^{({\rm cpx})}$ may be different from its vacuum value $\vev{W^{({\rm cpx})}}$.
When the deviation $W^{({\rm cpx})} - \vev{W^{({\rm cpx})}}$ is too large,
the last two terms of (\ref{eq:scalar-potential-ZandT-partialNoScaleCancl})
[i.e., the second term of (\ref{eq:scalar-potential-ZandT})] may no longer
be positive, and/or the K\"{a}hler moduli may start to decompactify.

To study the constraint from volume destabilization, one should use the 
full scalar potential (\ref{eq:scalar-potential-ZandT}, 
\ref{eq:scalar-potential-ZandT-partialNoScaleCancl}), but it is messy, 
complicated, and even worse, depends very much on the choice of 
Calabi--Yau geometry for compactification. In order to extract as robust 
lessons as possible, we begin with a little phenomenological approach instead. 
That is a) to ignore the first term of (\ref{eq:scalar-potential-ZandT}), 
and b) to deal with a limited number of parameters capturing all the 
influence of complex structure moduli evolution entering the remaining 
term---the second term---in (\ref{eq:scalar-potential-ZandT}). 
We bring in a bit of guess work in deriving the destabilization constraint, 
in order to overcome the disadvantage associated with (a).
It takes an extra effort to carry out model-specific analysis using the 
full scalar potential (without the guess work); discussion in 
section \ref{ssec:alg-nmb} can be regarded as a necessary first step.

There are only three different ways the field value of complex structure 
moduli $z$ enters into the second term of (\ref{eq:scalar-potential-ZandT}).
First, the total superpotential $W^{({\rm tot})}$ contains $W^{({\rm cpx})}(z)$ 
which depends on $z$, and secondly, we should also expect $z$-dependence 
in $a_i$'s of the racetrack superpotential. Although the overall factor 
$e^{{\cal K}^{({\rm cpx})}}$ also introduces $z$-dependence, this overall factor is more or less irrelevant to the 
study of instability toward decompactification 
${\rm Re}(T) \rightarrow \infty$. The first effect is parametrized by 
$\delta \widetilde{W}^{({\rm cpx})}_{\rm eff.}$ in 
\begin{align}
W^{({\rm tot})} =\frac{M_{\rm Pl}^3}{ [(4\pi){\rm Re}\vev{T}]^{3/2} } 
\left( \sum_i a_i N_i^2 e^{-\frac{2\pi}{N_i}T} \right)
  -\vev{W^{(T)}}
  +\frac{M_{\rm Pl}^3}{\sqrt{4\pi}} \delta \widetilde{W}_{\rm eff.}^{({\rm cpx})} ;
\label{eq:Wtot-parametrize4racetrack}
\end{align}
the value of $W^{({\rm cpx})}(z)$ is split into 
$W_0 + M_{\rm Pl}^3/\sqrt{4\pi} \times \delta \widetilde{W}^{({\rm cpx})}_{\rm eff.}$,
the vacuum value and deformation from it during the inflation and/or 
reheating process. 
We implement the other $z$-dependence by dealing with $a_2/a_1$ as 
another parameter (which may also be different from its vacuum value 
$\vev{a_2/a_1}$).

We study how much $\delta \widetilde{W}^{({\rm cpx})}_{\rm eff.}/a_1$ and $a_2/a_1$ 
can be changed without distorting the $T$-dependent potential 
$V_{\rm racetrack}(T)$ too much; the superpotential 
$W^{({\rm tot})}$ in (\ref{eq:Wtot-parametrize4racetrack}) is used for 
$V_{\rm racetrack}(T)$ instead of $W^{(T)}+W_0$, so that the potential 
$V_{\rm racetrack}(T)$ is now roughly equal to the second term 
of (\ref{eq:scalar-potential-ZandT}) with the irrelevant overall factor 
$e^{{\cal K}(z)}$ stripped off. 
As we have announced at the end section \ref{ssec:racetrack}, 
we choose the value $N_1 =68$, $N_2=67$ and $r = \vev{a_2/a_1} = - 1.05041$, 
so that $\vev{T}=25$ and the K\"{a}hler modulus is stabilized for the vacuum 
value of complex structure moduli $z = \vev{z}$.  
Although it is desirable to study the deformation of $V_{\rm racetrack}(T)$ for 
the complex two-dimensional parameter space of deformation, 
$(\delta \widetilde{W}^{({\rm cpx})}_{\rm eff.}/a_1 , a_2/a_1)$, we will carry out 
the study only along three real 1-parameter deformations in the parameter 
space in the following. \\

{\bf (i) variation of $W^{\rm (cpx)}(z)/a_1$}

\begin{figure}[tbp]
 \begin{center}
  \begin{tabular}{cc}
  \includegraphics[width=.5\linewidth]{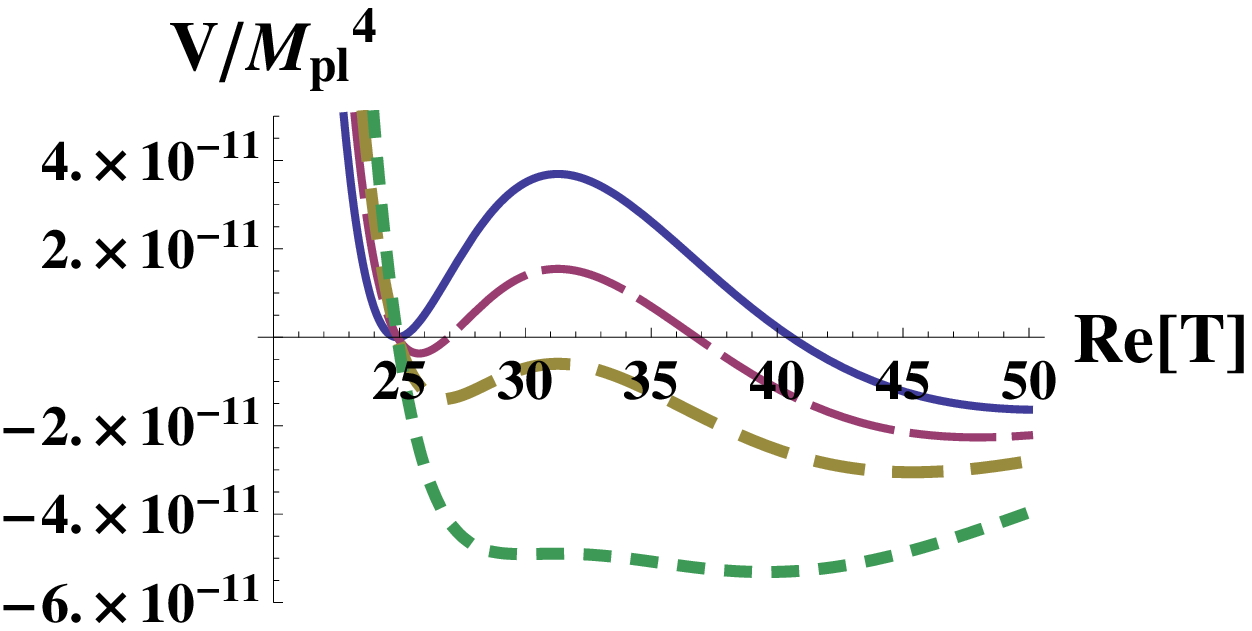} &  
  \includegraphics[width=.5\linewidth]{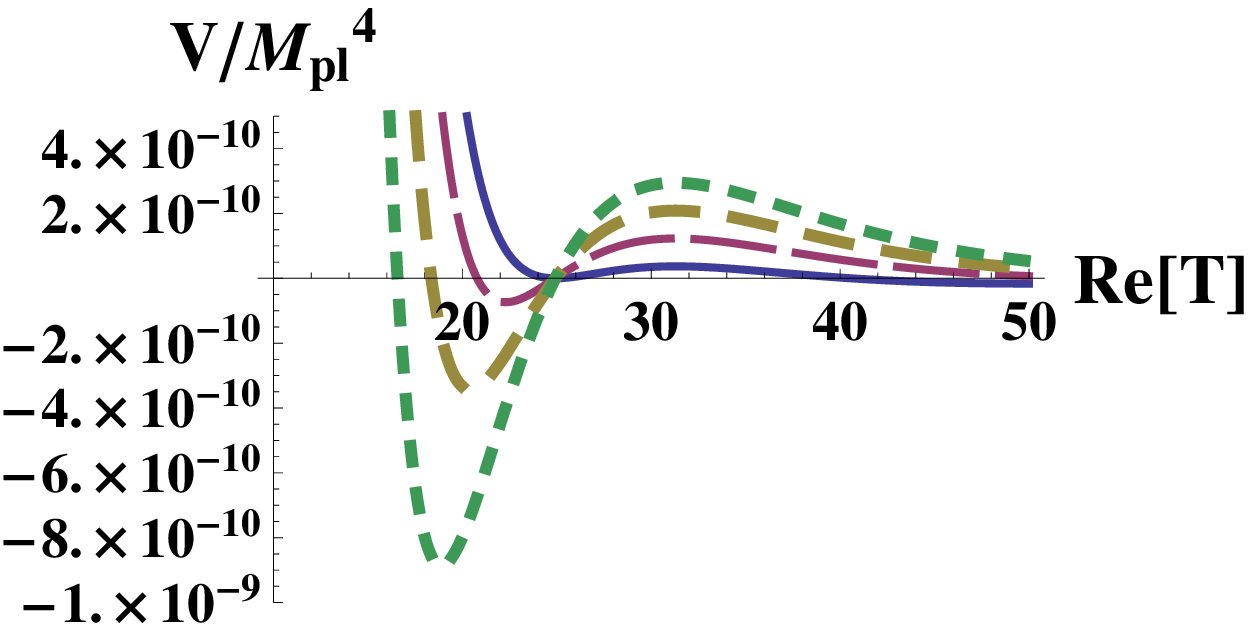}  \\
  (a) & (b) 
  \end{tabular}
  \caption{\label{fig:destabilization-Weff} 
The potential $V_{\rm racetrack}(T)$ with $W^{({\rm tot})}$ in 
(\ref{eq:Wtot-parametrize4racetrack}) for various values of 
$\delta \widetilde{W}^{({\rm cpx})}_{\rm eff.}(z)/a_1$; the vacuum parameters 
in (\ref{eq:vac-parameter}) are used here, 
and the other deformation parameter $a_2/a_1$ is held fixed at its vacuum 
value in this figure.
{\bf Panel (a):} the four curves from top to bottom (solid, long dashed, 
dashed and dotted) correspond to the deformation parameters 
$\delta \widetilde{W}^{({\rm cpx})}_{\rm eff}/a_1 =0$, 
$-0.0005$, $-0.001$ and $-0.002$, respectively. 
{\bf Panel (b):} the potential for the deformation 
$\delta \widetilde{W}^{({\rm cpx})}_{\rm eff}/a_1 = 0$,  
$0.002$, $0.004$ and $0.006$ are shown in the solid, long dashed, 
dashed and dotted curves, respectively.
}
 \end{center}
\end{figure}

Figure~\ref{fig:destabilization-Weff}~(a, b) shows how $V_{\rm racetrack}(T)$ 
changes as we change the value of 
$(\delta \widetilde{W}^{({\rm cpx})}_{\rm eff.})/a_1$ while keeping $a_2/a_1$ fixed 
at the vacuum value $r = \vev{a_2/a_1}$.
Starting from the vacuum value 
$\delta \widetilde{W}^{({\rm cpx})}_{\rm eff.}/a_1 =0$ 
(i.e., $W^{({\rm cpx})}(z) = W_0= - \vev{W}^{(T)} \simeq
 - 0.00122 \times a_1 M_{\rm Pl}^3$) and adding deformation 
$\delta \widetilde{W}^{({\rm cpx})}_{\rm eff.}$ in the negative real-valued region, 
we see that the barrier in $V_{\rm racetrack}(T)$ almost disappears by the time 
$\delta \widetilde{W}^{({\rm cpx})}_{\rm eff.}/a_1 \simeq -0.001$.  
Certainly the K\"{a}hler moduli dependence of the full scalar potential 
(\ref{eq:scalar-potential-ZandT}) is not the same as $V_{\rm racetrack}(T)$, 
because the first term provides uplift\footnote{Remember that 
the vacuum energy $V(z,T)$ vanishes at the vacuum value $z = \vev{z}$, 
$T = \vev{T}$ under the Kallosh--Linde tuning (\ref{eq:cond-KL-min}), and 
the potential energy must be positive for any small perturbation from 
$z = \vev{z}$ and $T = \vev{T}$.} at least around $T \sim \vev{T} = 25$.
With some uplift component added to the $T$-dependent potential in 
Figure~\ref{fig:destabilization-Weff}~(a), it is hard to imagine, however, 
that the uplift term drastically improves the potential for volume 
stabilization (even against quantum tunneling) for the range of deformation 
$\delta \widetilde{W}^{({\rm cpx})}_{\rm eff.}/a_1 \sim - 0.001$. 
Therefore, it will not be terribly bad to take this value as an estimate of 
the limit on how much $\delta \widetilde{W}^{({\rm cpx})}_{\rm eff.}$ can change 
without jeopardizing the K\"{a}hler moduli stabilization.

Perturbing the value of $\delta \widetilde{W}^{({\rm cpx})}_{\rm eff.}$ in the 
real positive region instead, we see in 
Figure~\ref{fig:destabilization-Weff}~(b) that the potential barrier of 
$V_{\rm racetrack}(T)$ becomes higher, but the local minimum around $T \sim 25$ 
gets deeper and deeper at the same time. 
When the value of $\delta \widetilde{W}^{({\rm cpx})}_{\rm eff.}/a_1$ is around 
$+ 0.004$, the potential dip in $V_{\rm racetrack}(T)$ is much more than the 
barrier height in their absolute values;  
the uplifting contribution from the first term 
of (\ref{eq:scalar-potential-ZandT}) will push this dip at least above the zero 
energy density of $V(z,T)$. This means that the $T$-dependence of the full 
potential $V(z,T)$ in (\ref{eq:scalar-potential-ZandT}) is different from 
that of $V_{\rm racetrack}(T)$ considerably, and we cannot say the K\"{a}hler 
moduli $T$ remains to be stabilized without studying the full potential 
(\ref{eq:scalar-potential-ZandT}) for such a value of 
$\delta \widetilde{W}^{({\rm cpx})}_{\rm eff.}$. This argument does not rule out a possibility 
that the first term of (\ref{eq:scalar-potential-ZandT}) 
provides just enough uplift around $T \sim 20$ so that $\rho = V(T,z)$ 
becomes barely positive, but not as high as the potential barrier around 
$T \sim 30$; we do not have a concrete idea of how to coordinate the first 
and second terms in (\ref{eq:scalar-potential-ZandT}) for that to happen, 
however.

For these reasons, and in this meaning, we see that the volume stabilization 
in $V_{\rm racetrack}(T)$ remains to be reliable as long as the value of 
$W^{({\rm cpx})}(z)$ differs from its vacuum value $W_0$ within the range 
\begin{align}
- 0.001 \lesssim \delta \widetilde{W}^{({\rm cpx})}_{\rm eff.}/a_1 \lesssim 0.004.
\label{eq:Wtil-cpx-eff-limit}
\end{align}
This estimate of the limit can be read as that of the deformation in  
$a_1^{-1} \int_X G \wedge \Omega = \sum_a (n^{R}_a - \tau n^{NS}_a) \Pi_a/a_1$, 
since we chose the normalization of $\delta \widetilde{W}^{({\rm cpx})}_{\rm eff.}$ 
that way in (\ref{eq:Wtot-parametrize4racetrack}).\\

{\bf (ii) variation of $a_2/a_1(z)$}

\begin{figure}[tbp]
 \begin{center}
  \begin{tabular}{cc}
  \includegraphics[width=.5\linewidth]{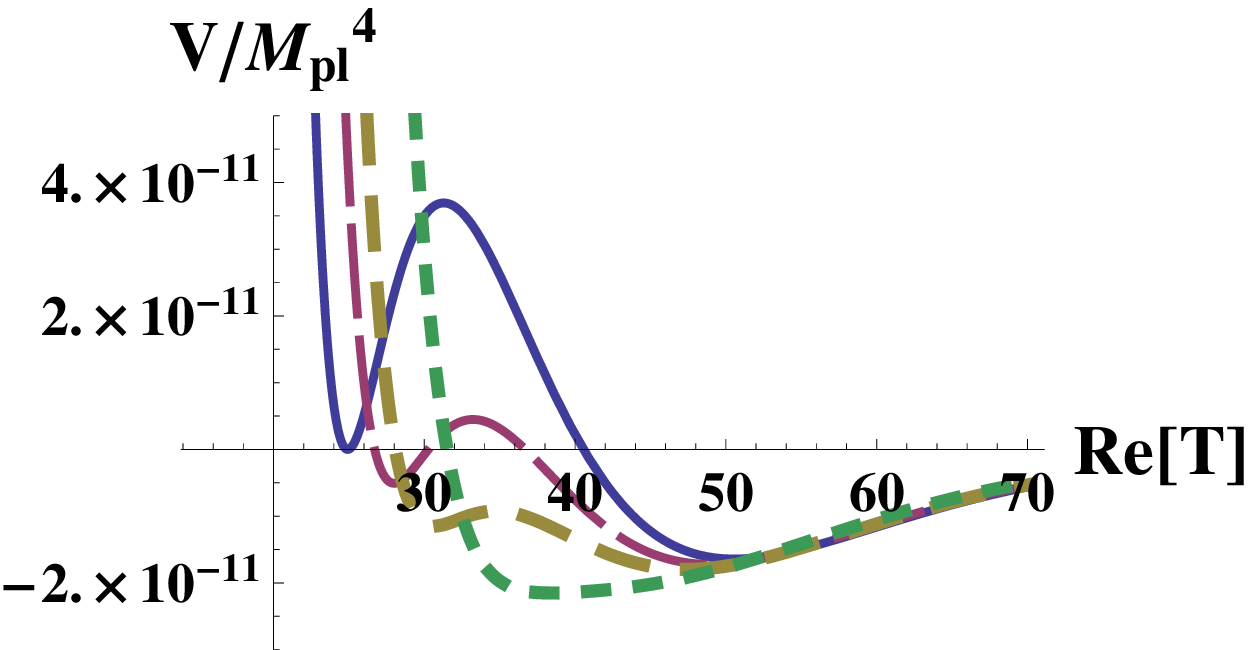} &  
  \includegraphics[width=.5\linewidth]{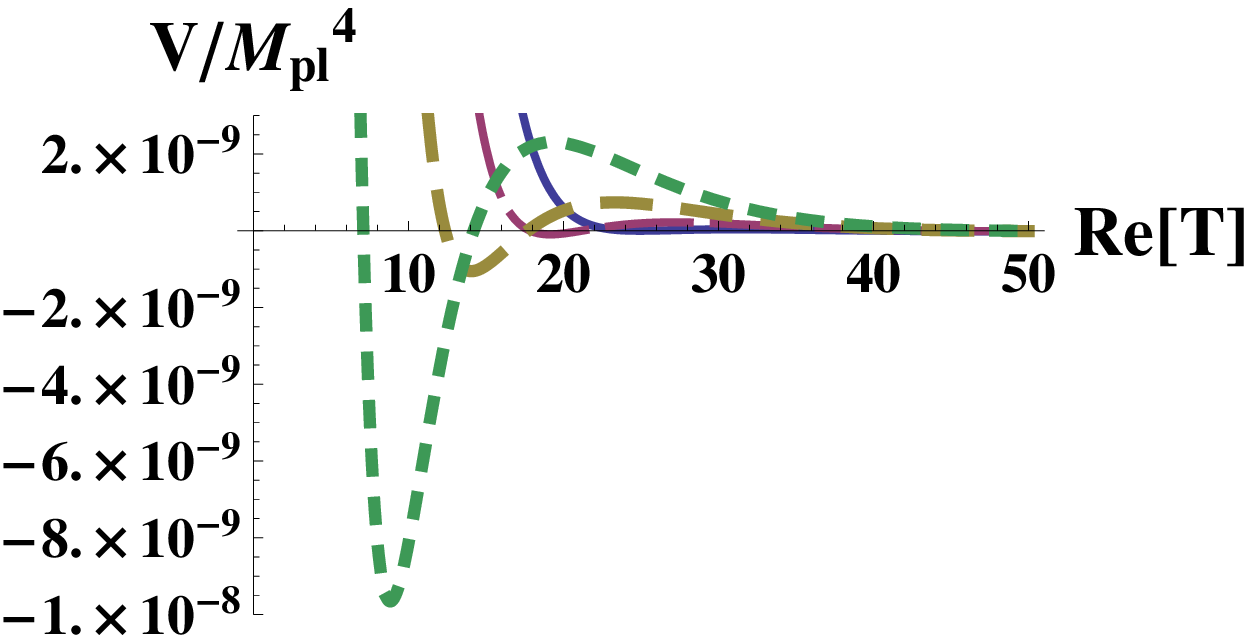}  \\
  (a) & (b) 
  \end{tabular}
  \caption{\label{fig:destabilization-a2a1} 
The potential $V_{\rm racetrack}(T)$ with $W^{({\rm tot})}$ in 
(\ref{eq:Wtot-parametrize4racetrack}) for various values of $a_2/a_1$; 
the value of $W^{({\rm cpx})}(z)$ is fixed at $W_0$ in this figure. 
{\bf Panel (a):} the four curves (solid, long dashed, dashed and dotted) are 
for $a_2/a_1 \simeq - 1.0504$, $-1.0530$, $-1.0550$ and $-1.0600$, 
respectively.
{\bf Panel (b):} the parameter value is set at 
$a_2/a_1 = -1.0504$, $-1.0450$, $-1.0400$ and $-1.0350$ in the curves drawn 
in the solid, long dashed, dashed and dotted lines, respectively.
The potential $V_{\rm racetrack}(T)$ drawn in the solid line in (a) and (b) are 
the same, the one at the vacuum $a_2/a_1 = \vev{a_2/a_1}$.}
 \end{center}
\end{figure}

Figure~\ref{fig:destabilization-a2a1} shows how $V_{\rm racetrack}(T)$ changes 
when the value of $a_2/a_1$ is different from its vacuum value 
$r = \vev{a_2/a_1} \simeq -1.05041$ instead, while the value of 
$W^{({\rm cpx})}(z)$ somehow remains to be $W_0$.
We can use the numerical results in the figure, to set a limit 
\begin{equation}
-1.0550 \lesssim a_2/a_1 \lesssim -1.0400
\label{eq:limit-a2-a1-ratio-result}
\end{equation}
for the same reason as in the analysis of changing 
$\delta \widetilde{W}^{({\rm cpx})}_{\rm eff.}/a_1$. 
The potential $V_{\rm racetrack}(T)$ is highly sensitive to the value of 
$a_2/a_1$, when the value of $W^{({\rm cpx})}(z)/a_1$ is held fixed, 
and there is not much room around the vacuum value 
$\vev{a_2/a_1} \simeq -1.05041$. This result is not hard to imagine 
from the discussion in section \ref{ssec:racetrack}, because of the 
high sensitivity of the potential $V_{\rm racetrack}(T)$ on the vacuum value 
$\vev{a_2/a_1}$. If we replace the vacuum parameters 
in (\ref{eq:vac-parameter}) by more negative $r = \vev{a_2/a_1}$ and smaller 
$N_1$, we expect larger range of deformation in $a_2/a_1$ from the new vacuum 
value will be allowed, than in (\ref{eq:limit-a2-a1-ratio-result}); 
this comes at the cost of limiting the energy density during inflation 
from above, however. \\

{\bf (iii) variation of both $\delta \widetilde{W}^{({\rm cpx})}_{\rm eff.}/a_1$ and $a_2/a_1(z)$}

We have so far searched only along two real-valued 1-parameter deformations 
in the phenomenological parameter space 
$(\delta \widetilde{W}^{({\rm cpx})}_{\rm eff.}/a_1, a_2/a_1)$ of 
$V_{\rm racetrack}(T)$. This leaves a room for some combination of changes 
in $W^{({\rm cpx})}(z)$ and in $a_2/a_1$ so that the potential barrier remains 
in $V_{\rm racetrack}(T)$. There is no obvious strategy to 
look for such a coordinated changes, or to claim their absence. We just 
try one more, a 1-parameter deformation in $W^{({\rm cpx})}$ and 
$a_2/a_1$ so that $V_{\rm racetrack}(T)$ remains to have vanishing energy 
at the local minimum around $T \sim \vev{T} \simeq 25$ (the energy density 
of the full potential (\ref{eq:scalar-potential-ZandT}) is positive); 
this corresponds to focus on a subspace of 
$(\delta \widetilde{W}^{({\rm cpx})}_{\rm eff.}/a_1, a_2/a_1)$ satisfying a 
relation 
\begin{equation}
   - \frac{M_{\rm Pl}^3 \; a_1 N_1}{(4\pi {\rm Re}\vev{T})^{3/2}} 
        \left(- \frac{a_1}{a_2} \frac{N_1}{N_2} \right)^{N_2}
    =W^{\rm (cpx)}
    =W_0 +\frac{M_{\rm Pl}^3}{\sqrt{4\pi}}
       \delta \widetilde{W}^{({\rm cpx})}_{\rm eff.}, 
  \label{eq:conspire-KL}
\end{equation}
which generalizes (\ref{eq:W0-def}).

As in Figure~\ref{fig:destabilization-conspire}~(a),
for the choice of $a_2/a_1$ closer to zero than the vacuum value 
$\vev{a_2/a_1} \simeq -1.0504$ [i.e., 
$\delta \widetilde{W}^{({\rm cpx})}_{\rm eff.}/a_1 \leq 0$], the potential barrier 
$V_{\rm racetrack}(T)$ becomes higher (read the caption carefully), but 
the second minimum (at larger value of $T$) also gets deeper more rapidly. 
The K\"{a}hler moduli field $T$ may remain marginally stable 
in $V_{\rm racetrack}(T)$ for $a_2/a_1 = -1.035$, but it will not 
be against quantum tunneling for a value even closer to zero. Thus, this destabilization argument sets 
a limit in the deformation satisfying (\ref{eq:conspire-KL}) as follows:
\begin{figure}[tbp]
 \begin{center}
  \begin{tabular}{cc}
  \includegraphics[width=.5\linewidth]{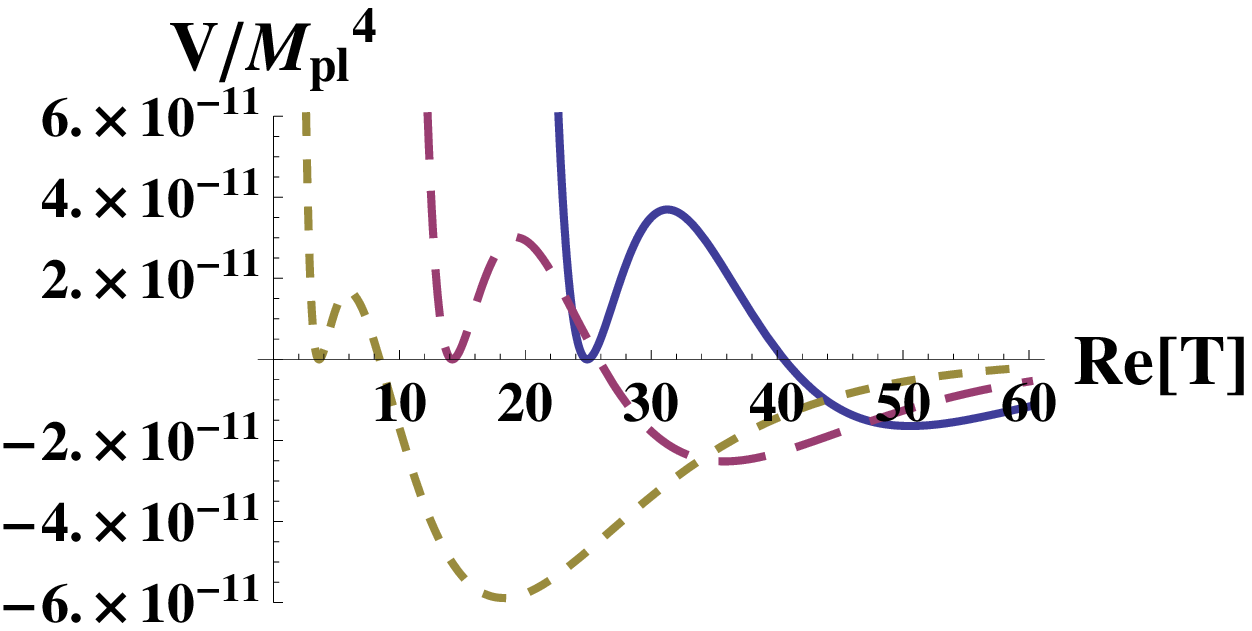} &  
  \includegraphics[width=.5\linewidth]{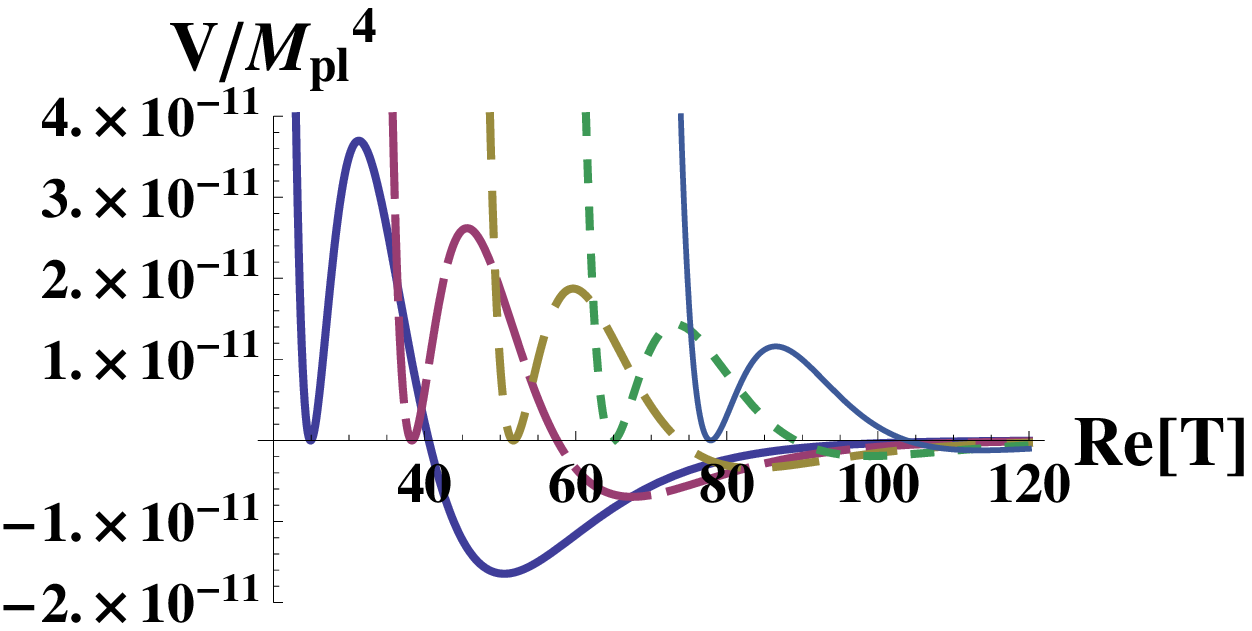}  \\
  (a) & (b) 
  \end{tabular}
  \caption{\label{fig:destabilization-conspire} 
The potential $V_{\rm racetrack}(T)$ with $W^{({\rm tot})}$ in 
(\ref{eq:Wtot-parametrize4racetrack}) with different sets of deformation 
parameters $(\delta\widetilde{W}^{({\rm cpx})}_{\rm eff.}/a_1, a_2/a_1)$ 
satisfying the relation (\ref{eq:conspire-KL}).
{\bf Panel (a):} $V_{\rm racetrack}(T)$ is drawn for the deformation parameters  
$(0, -1.0504)$, $(-0.007, -1.0350)$ and $(-0.027, -1.0200)$ in 
the solid, dashed and dotted line, respectively, after multiplied by 
$1$, $10^{-1}$ and $10^{-2}$, respectively; this means, for example, 
that the local maximum of the potential for $(-0.027, -1.0200)$ 
is about $2 \times 10^{-9} \times M_{\rm Pl}^4$. 
{\bf Panel (b):} the solid, long dashed, dashed, dotted and thin solid lines 
show the potential $V_{\rm racetrack}(T)$ for the sets of deformation parameters  
$(0, -1.0504)$, $(0.0031, -1.07)$, 
$(0.0040, -1.09)$, $(0.0042, -1.11)$ and 
$(0.0043, -1.13)$, respectively, multiplied by a factor 
$1$, $10$, $10^2$, $10^3$ and $10^4$, respectively; this means, for example, 
that the local maximum of $V_{\rm racetrack}(T)$ is about 
$10^{-15} \times M_{\rm Pl}^4$ for the deformation $(0.0043, -1.13)$. 
}
 \end{center}
\end{figure}
%
\begin{equation}
  -0.007 \lesssim \delta \widetilde{W}^{({\rm cpx})}_{\rm eff.}/a_1, \qquad 
  a_2/a_1 \lesssim -1.035.
\end{equation}
With the coordinated deformation in (\ref{eq:conspire-KL}), certainly the 
destabilization limit above has been relaxed a bit 
from (\ref{eq:Wtil-cpx-eff-limit}, \ref{eq:limit-a2-a1-ratio-result}), 
but not much. 
For deformation in the other direction, however, the second minimum in 
the potential $V_{\rm racetrack}(T)$ becomes less and less pronounced for even 
more negative value of $a_2/a_1$ relatively to the vacuum value $-1.0504$, 
as one can see in Figure~\ref{fig:destabilization-conspire}~(b). 
There may be no danger of destabilization for such a change, and the lower 
bound on $a_2/a_1$ disappears. This considerably relaxes the constraint 
$-1.0550 \lesssim a_2/a_1$ obtained earlier (only the $W^{({\rm cpx})} \leq 0$ 
region is probed under the relation (\ref{eq:conspire-KL})). 
The potential barrier height is also reduced considerably 
at the same time, however, for such a deformation 
in $(\delta \widetilde{W}^{({\rm cpx})}/a_1, a_2/a_1)$; see the caption of the figure. In order to support 
inflation at high energy scale, that may not be appropriate.

\section{Constraint on Complex Structure Deformation during/after Inflation}
\label{sec:mirror_quintic}

In the previous section, we have studied how much complex structure moduli 
can be deformed from their vacuum values during the inflation/reheating 
process by (I) using parameters representing the effects of complex structure 
deformation, and (II) studying the potential $V_{\rm racetrack}(T)$ (rather than 
the full scalar potential (\ref{eq:scalar-potential-ZandT})). 
In sections \ref{ssec:first} and \ref{ssec:alg-nmb}, we make it clear 
what the constraint on the deformation means in Type IIB Calabi--Yau 
orientifold compactifications, and at the same time, we carry out 
necessary technical preparation so that the analysis using the full 
scalar potential is possible. In section \ref{ssec:solution}, 
a few ideas of how to avoid/satisfy the constraint are discussed. 

\subsection{An Attempt of Implementing KYY in Type IIB Orientifolds}
\label{ssec:first}

\subsubsection{A Brief Note on Mirror-Quintic}
\label{sssec:review-m.quintic}

We use the ``mirror-quintic'' as a Calabi--Yau 3-fold $X$ for Type IIB 
orientifold compactification and something similar to the mirror quintic, 
for the presentation purpose in sections \ref{ssec:first} 
and \ref{ssec:alg-nmb}; much of the statements will remain valid 
for broader class of Type IIB Calabi--Yau orientifolds and for F-theory 
compactifications, however. To be more precise, we use the superpotential 
$W^{({\rm cpx})} = W_{GVW}$ and 
$K^{({\rm cpx})}/M_{\rm Pl}^2 = - \ln [ i \int_X \Omega \wedge \overline{\Omega} ]$ 
based on the mirror-quintic, while we still continue to use only one 
K\"{a}hler modulus chiral multiplet $T$ and the no-scale type K\"{a}hler 
potential (see footnote~\ref{fn:multi-h11}). Obviously the statements 
in sections \ref{ssec:first} and \ref{ssec:alg-nmb} should not be taken  
as precise, rigorous results about compactifications on the mirror 
quintic $X$; the following discussion is rather meant to be a test study 
of general properties in Type IIB or F-theory compactifications.  

Before getting started, we leave a brief summary of known facts about 
the mirror-quintic that are used in the following discussion.

The mirror-quintic Calabi--Yau 3-fold $X$ has only one complex structure 
modulus. It is regarded as a crepant resolution of 
$\Z_5 \times \Z_5 \times \Z_5$ orbifold of a geometry given by 
\begin{equation}
 \left\{ [X_1:\cdots : X_5] \in \P^4 \; | \; 
     \sum_{i=1}^5 (X_i)^5 - 5 \psi \prod_{i=1}^5 X_i = 0  \right\}.
\end{equation} 
The complex structure modulus of the mirror-quintic is parametrized 
best by $\zeta$ (or $z_\psi$), where 
\begin{equation}
  \zeta := \frac{1}{2\pi i} \ln (z_\psi), \qquad z_\psi := \frac{1}{(5\psi)^5}.  
\end{equation}
In such a Type IIB set-up, the two moduli fields $(\tau, \zeta)$ correspond 
to what we loosely referred to as ``complex structure moduli $z$'' in 
earlier sections. The period integral is expressed for a symplectic basis 
of $H_3(X; \Z)$ as follows \cite{Candelas:1991}
\begin{equation}
  \Pi = 
 \left(\begin{array}{c} 
   \Pi_1 \\ \Pi_2 \\ \Pi_3 \\ \Pi_4 \end{array} \right)  = 
\left( \begin{array}{r} 
   1 \\ \zeta \\
   - \frac{5}{2} \zeta^2 - \frac{11}{2} \zeta + \frac{25}{12} \\
   \frac{5}{6} \zeta^3 + \frac{25}{12} \zeta
             - i \frac{\chi(X) \zeta_3}{(2\pi)^3} 
   \end{array} \right) + {\cal O}(e^{2\pi i n \zeta})_{n \geq 1},  
\label{eq:period-integral}
\end{equation}
where $\zeta_3$ is meant to be $\zeta(3) = 1.202 \cdots$, and $\chi(X)=200$ 
is the topological Euler number of the mirror-quintic Calabi--Yau 3-fold $X$.
The K\"{a}hler potential of the chiral multiplet $\zeta$ is given by 
\begin{align}
K^{({\rm cpx})}/M_{\rm Pl}^2
&= -\ln \left( i \int_X \Omega \wedge \overline{\Omega} \right) = 
 - \ln \left( - i \Pi^\dagger \cdot \Sigma \cdot \Pi \right) \notag \\
&= - \ln \left( \frac{5}{6} [(\zeta - \bar{\zeta})/i]^3 
     + \frac{50}{\pi^3} \zeta_3
     + {\cal O}(e^{- 2\pi {\rm Im}(\zeta)})   \right)
\label{eq:Kahler-m.quintic}
\end{align}

Changing the phase of the parameter $z_\psi$ by $2\pi$, we come back to the 
same point in the complex structure moduli space of the mirror-quintic. 
This results in a shift 
\begin{equation}
\zeta \longrightarrow \zeta' = \zeta + 1,
\label{eq:shift}
\end{equation}
but the K\"{a}hler potential above remains invariant. 
The period integral (\ref{eq:period-integral}) does change 
under the shift (\ref{eq:shift}), but the change is in the form of 
\begin{align}
\Pi (\zeta) \longrightarrow 
\Pi (\zeta +1) =
M_{\infty} \cdot \Pi (\zeta),
\quad M_\infty 
:=\left(\begin{array}{cccc} 
    1& 0& 0& 0 \\ 1& 1& 0& 0 \\ -8& -5& 1& 0 \\ 5& -3& -1& 1
\end{array}\right),
\end{align}
and is regarded as monodromy transformation of a integral symplectic 
basis of $H_3(X; \Z)$.
This discrete shift symmetry may be regarded as a continuous shift symmetry 
(in the K\"{a}hler potential) approximately in the large complex structure 
region, ${\rm Im}(\zeta) \gg 1$. 

The complex structure for $\zeta$ and for $\zeta' = \zeta + 1$ are regarded 
the same physically, only in the absence of 3-form fluxes in Type IIB 
compactification. 3-form fluxes in Type IIB, $G = F^{(3)} - \tau H^{(3)}$, 
are characterized by the choice of integers $F^{(3)} \leftrightarrow \{n^R_a\}$
and $H^{(3)} \leftrightarrow \{ n^{NS}_a \}$. For a given choice of fluxes, 
$\{n^R_a\}$ and $\{n^{NS}_a\}$, two choices of complex structure, $\zeta$ and 
$\zeta'$ are not equivalent in physics, because 
$\sum_a (n^R_a -\tau n^{NS}_a)\Pi_a(\zeta)$ is not the same as that 
for $\Pi_a(\zeta')$. One could get the period integral $\Pi_a(\zeta'_a)$
back to $\Pi_a(\zeta)$ by the monodromy transformation, but the flux quanta 
in the new symplectic integral basis $\{n^R_a\}$ and $\{n^{NS}_a\}$ are not 
the same as before. Due to this monodromy mechanism on the moduli 
space of complex structure moduli, its covering space is a more appropriate 
moduli space for physics in the presence of flux on $X$. 

It is possible that there are correction terms to the K\"{a}hler potential 
(\ref{eq:Kahler-m.quintic}) generated in the presence of fluxes. However, 
corrections will be small, relatively by 
$\alpha' G^2 \sim 1/\vev{\omega}^3 \ll 1$ in the regime of our interest 
in this article (${\rm Re}\vev{T} \approx {\cal O}(10)$ or somewhat larger). 
We do not pay attention to the correction terms in the K\"{a}hler potential 
for this reason. 

All the properties described so far are not specific to the mirror-quintic 
Calabi--Yau 3-fold, but hold true for most of Calabi--Yau 3-folds. 
``{\it All the properties}'' include the approximate shift symmetry in the K\"{a}hler 
potential in the large complex structure moduli region in the absence of flux, 
the monodromy group action on the moduli space and the flux quanta, 
rational number coefficients in the period integral (like (\ref{eq:period-integral})) 
determined by topology of $X$ as well as the mirror geometry 
of $X$ \cite{CICY}, and the small corrections due to flux controlled by 
$1/[{\rm Re}\vev{T}]^{3/2}$. 

Such an idea of exploiting an approximate shift symmetry and monodromy 
in the complex structure moduli for inflation in Type IIB/F-theory has been 
presented in \cite{cpxstrinflation,Silverstein, Hebecker}; this is a mirror version 
of the same set of ideas exploited in axion monodromy 
inflation \cite{axion-monodromy}.
This set-up also shares with D3--D7 inflation an approximate continuous 
shift symmetry in the K\"{a}hler potential that is broken by flux or other 
effects in the superpotential \cite{D3D7-shift}. 
In this sense, the use of complex structure moduli for string inflation 
is another variation of the same theme that has been pursued for the last 
decade. 

\subsubsection{Imaginary Part of $\zeta$ as an Inflaton?}

In the large complex structure region ${\rm Im}(\zeta) \gg 1$, 
the kinetic term is approximately 
\begin{equation}
 {\cal L} \simeq \frac{3 M_{\rm Pl}^2}{(\zeta- \bar{\zeta})^2} 
    |\partial \zeta|^2.
\end{equation}
This means that it is better to parametrize the complex $\zeta$-plane 
by two real fields $\sigma$ and $\varphi$ as 
\begin{equation}
  \zeta = i e^{2\sigma}(1 + i 2 \varphi), 
\end{equation}
so that the kinetic term is close to the canonical one,
\begin{equation}
 {\cal L} \simeq - 3 M_{\rm Pl}^2 \left[ (\partial \sigma)^2 + 
  \left( \partial \varphi + 2 (\partial \sigma) \varphi \right)^2 \right].
\end{equation}

Because the superpotential $W_{GVW}$ for the period integral in 
(\ref{eq:period-integral}) is approximately a polynomial in $\zeta$ 
that is at most cubic, it is a combination of exponential functions 
in $\sigma$, and  polynomial in $\varphi$. It is better to 
expect the slow-roll evolution along the $\varphi$ direction, at least 
relatively to the $\sigma$ direction. If we want the $\varphi$ field 
to evolve by about ${\cal O}(10)$--${\cal O}(100)$ in order to earn 
sufficiently large e-fold number, then it means under the assumption of 
large complex structure region, $\sigma \gtrsim 1$, that the real 
part of $\zeta$ changes by of order $e^{2\sigma} \times (10\mbox{--}100)$.

\subsubsection{A KYY Look-alike in Type IIB Orientifolds}
\label{sssec:KYY-IIB}

If the real part of $\zeta$ were to change by $e^{2\sigma} \times 
{\cal O}(10) \gg 1$ in order to drive slow-roll inflation, the value of 
$\tilde{W}^{({\rm cpx})}_{\rm eff.}$ would also change by much more than a value of 
order unity. We have seen in section \ref{ssec:phen-limit-destabil}, 
however, that the scalar potential stabilizing the K\"{a}hler modulus $T$ is 
distorted so much then that we are no longer confident whether $T$ remains to be 
stabilized.  It is thus possible to conclude that slow-roll inflation may 
by driven purely by the K\"{a}hler moduli field $T$, not by 
$z = (\zeta, \tau)$; 
right-handed sneutrino scenario may be relevant only through the kinetic/mass 
mixing between $\delta T$ and $\delta z$ around the supersymmetric minimum 
reached at the end of inflation, a possibility that we have already mentioned 
in section \ref{sec:KL-problem}. 

This constraint on the case for inflation in complex structure moduli has 
its origin in string compactification, because the limits on the variation in 
$W^{({\rm cpx})}$ came from stabilization of K\"{a}hler moduli. 
A similar limitation on large-field inflation model was encountered before, 
however, in the context of supergravity, rather than in string theory. 
That was an observation that it is not easy to implement chaotic inflation 
into 4D ${\cal N}=1$ supergravity; starting with a simple monomial 
superpotential $W \sim \Phi^n$, we obtain a potential $|\Phi^{n-1}|^2$ from 
the F-term in rigid supersymmetry, but the $-3|W/M_{\rm Pl}|^2$ term 
in the 4D supergravity scalar potential would become more important 
for super-Planckian $\Phi$ than the F-term. The potential not only stops 
being a monomial, but also even becomes negative. An idea to get around 
the problem in supergravity by Kawasaki, Yamaguchi and Yanagida (KYY) \cite{KYY} was to consider an effective 
superpotential of the form $W = X \; {\rm fcn}_1(\Phi)$ and a K\"{a}hler 
potential $K = X^\dagger X + \cdots + {\rm fcn}_2(\Phi + \Phi^\dagger)$ so that 
the value of $X$ remains zero during inflation, and so do the values of 
$W$ and $-3|W/M_{\rm Pl}|^2$ consequently.

It appears that it is possible to implement such a situation  
by choosing the 3-form fluxes.\footnote{See \cite{KYY-alike} for other 
attempts of implementing the idea of \cite{KYY} in string theory.}\raisebox{5pt}{,}\footnote{In the end, it turns out that this implementation does not work. 
Busy readers can proceed to section \ref{ssec:alg-nmb}.} 
If we choose $n^R_1  \neq 0$, $n^R_{2,3,4} = 0$ 
and $n_a^{NS}$'s generic in the mirror-quintic $X$, then 
\begin{equation}
 W^{({\rm cpx})} = \frac{M_{\rm Pl}^3}{\sqrt{4\pi}} 
       \left(n^R_1 - \tau \left(\sum_a n^{NS}_a \Pi_a(\zeta)\right) \right).
\label{eq:KYY-lookalike-IIB}
\end{equation}
Thus, if it is possible to choose $n_1^R$ so that the tuning of Kallosh-Linde (\ref{eq:cond-KL-min}) is achieved,  
\begin{equation}
\frac{M_{\rm Pl}^3}{\sqrt{(4\pi)}} n^R_1 + \vev{W^{(T)}} = 0,
\label{eq:KL-tuning-RRflux-gaugino}
\end{equation}
then 
the remaining terms in the superpotential 
$W^{({\rm cpx})}(\zeta,\tau) - \vev{W^{({\rm cpx})}}$ is of the form introduced in 
\cite{KYY}; $\tau$ is for $X$ and $\sum_a n^{NS}_a \Pi_a(\zeta)$ for 
${\rm fcn}_1(\Phi)$. 
Can this combination remain very small during inflation, just like 
in \cite{KYY}, so that the problem of volume destabilization in string 
theory is also avoided?

The idea looks nice, but the tuning 
condition (\ref{eq:KL-tuning-RRflux-gaugino}) can never be satisfied. 
The Ramond--Ramond 3-form flux quantum $n_1^R$ is an integer, but the value of 
$\vev{W^{(T)}}$ is orders of magnitude smaller than $M_{\rm Pl}^3$ as long 
as we work in the moderate large radius regime. If the Kallosh--Linde 
tuning (\ref{eq:cond-KL-min}) is to be achieved, that should be 
done not in a way as naive as in (\ref{eq:KL-tuning-RRflux-gaugino}).
Because of the value  $\vev{W^{(T)}} \ll M_{\rm Pl}^3/\sqrt{4\pi}$, the vacuum vale 
$\vev{W_{GVW}}$ should vanish in some approximation; 
either small corrections in the approximation scheme or contributions 
to the superpotential $W^{({\rm tot})}$ other than $W_{GVW}$ may achieve 
the tuning against $W^{(T)}$..

One can think of setting $n_1^R = 0$ and assume that there is a term 
other than $W^{(T)}$ and $W_{GVW}$ in the superpotential $W^{({\rm tot})}$ whose 
vacuum value is $W_0$. Such a contribution to $W^{({\rm tot})}$ may come, in 
principle, from condensation of operators in a gauge theory supported on 
D-branes in $X$. Thus, the tuning (\ref{eq:cond-KL-min}) may still be achieved. 
In this case, one should note that the Ramond--Ramond 3-form flux is 
completely absent. The F-term potentials of $\zeta$ and $\tau$ give rise to 
\begin{eqnarray}
 V(T, \zeta, \tau) & \supset & 
   \frac{1}{2{\rm Im}(\tau)} \frac{1}{[2{\rm Im}(\zeta)]^3}
   \frac{1}{ [ {\rm Re}(T)]^3 } \frac{M_{\rm Pl}^4}{4\pi} \nonumber \\
   & & \qquad \qquad \times |\tau|^2 \times 
    \left\{ |f_N(\zeta)|^2
         + 3|f_N(\zeta) - (1/3) (\zeta-\bar{\zeta})\partial_\zeta f_N(\zeta)|^2
    \right\}, 
\end{eqnarray}
driving ${\rm Im}(\tau)$ to zero, similarly to the model of \cite{KYY}.
Here, 
\begin{equation}
  f_N(\zeta) := \sum_a n^{NS}_a \Pi_a(\zeta).
\end{equation}
The current set-up is different from that in \cite{KYY}, however, in that 
the K\"{a}hler potential of $\tau$ is singular at $\tau = 0$ (remember that $K/M_{\rm Pl}^2 \supset -\ln [(\tau-\bar{\tau})/i]$ ), whereas 
Ref. \cite{KYY} assumes that the target space $(X, \Phi)$ is smooth at $X = 0$.
The singularity of the K\"{a}hler potential at $\tau = 0$ and a possible 
$\zeta$-dependence of the coefficients $a_i$'s in $W^{(T)}$  results in 
such terms as 
\begin{equation}
 V(T, \zeta, \tau) \supset \frac{1}{{\rm Im}(\tau)}
       \frac{|\partial_\zeta W^{(T)}|^2}{M_{\rm Pl}^2}.
\end{equation}
Eventually a minimum may be formed around 
${\rm Im}(\tau) \sim W^{(T)}/[M_{\rm Pl}^3 f_N(\zeta)]$. 
If all things work properly in this way, and if the field value of $\tau$ 
keeps track of this minimum while the value of $\zeta$ changes over time, 
then the combination $M_{\rm Pl}^3 \tau f_N(\zeta) \sim W_{GVW}$ remains of 
order $\vev{W^{(T)}}$. This observation brings a hope that the volume 
stabilization constraint in section \ref{ssec:phen-limit-destabil} may 
actually be satisfied. 

The crude argument above does not pay close enough attention to 
the power counting of ${\rm Im}(\zeta)$ or ${\rm Re}(T)$, whose value we assume to be 
somewhat larger than 1. Given the fact that the constraints on 
$W^{({\rm cpx})}$---both the upper bound and lower bound---come 
at the same order as its vacuum value 
$W_0=-\vev{W^{(T)}}$, the power counting of 
${\rm Re}(T)$ and ${\rm Im}(\zeta)$ should be a crucial step to see 
whether the Type IIB implementation [(\ref{eq:KYY-lookalike-IIB}) with 
$n_1^R = 0$] of \cite{KYY} works or not.  
We did not choose to carry out such a careful study because we are referring to 
a potential $V(T, \zeta, \tau)$ in the ${\rm Im}(\tau) \ll 1$ region, 
and small ${\rm Im}(\tau) = e^{-\phi}$ implies large string coupling. 
The K\"{a}hler potential will receive corrections to the one 
$K \propto - \ln [(\tau-\bar{\tau})/i]$ in the strong coupling $e^{\phi} \gg 1$ 
regime, and it is hard to see to what extent such an analysis in 
perturbative Type IIB string theory is reliable. 

If we are to seek for a set up that is ${\rm SL}(2; \Z)$-equivalent to 
the flux superpotential (\ref{eq:KYY-lookalike-IIB}) with $n_1^R = 0$, 
that is in the form of 
\begin{equation}
 W_{GVW} = M_{\rm Pl}^3 \frac{1}{\sqrt{4\pi}} 
  \left(n^R - \tau n^{NS}\right) \cdot f(\zeta), 
\label{eq:KYY-lookalike-IIB-eqv}
\end{equation}
with $n^R$ and $n^{NS}$ both integers. Repeating the same analysis as above, 
however, we see that the valley in the scalar potential $V(\zeta, \tau, T)$ 
is in 
\begin{equation}
  n^R - \tau n^{NS} \approx
     \frac{W^{(T)} \sqrt{4\pi}/M_{\rm Pl}^3}{\partial_\zeta f(\zeta)};
\end{equation}
the combination $(n^R-\tau n^{NS})$ remains small along the potential valley, 
as in \cite{KYY}; the problem is that ${\rm Im}(\tau)$ also remains very 
small along the valley, and the string coupling large, provided $n^{NS} \neq 0$.
Thus, such an ${\rm SL}(2; \Z)$-equivalent description is still unreliable. 

The only ${\rm SL}(2; \Z)$-equivalent description 
of (\ref{eq:KYY-lookalike-IIB}) with $e^\phi \ll 1$ is the one with $n^{NS}=0$. This time, 
only the Ramond--Ramond fluxes are present, 
and the NS--NS flux is completely absent. This then suggests that the dilaton 
cannot be stabilized by the Gukov--Vafa--Witten superpotential. Indeed, 
one can see by writing down the scalar potential of the total system 
that the potential has a runaway direction 
${\rm Im}(\tau) \longrightarrow \infty$; although the total system includes 
additional terms (such as gaugino condensations) other than $W_{GVW}$, 
it is sufficient for justification of the runaway-claim above to assume 
that the Kallosh--Linde tuning (\ref{eq:cond-KL-min}) is achieved in a 
stabilized vacuum.

If there were a stabilized minimum reached after inflation 
in the description (\ref{eq:KYY-lookalike-IIB}) with $n_1^R=0$ 
or (\ref{eq:KYY-lookalike-IIB-eqv}), then it should be possible to map
the vacuum into the weak string coupling region by the ${\rm SL}(2;\Z)$ 
transformation. That fact that we can find the runaway situation at best 
implies that there is no stable minimum in the two earlier descriptions, 
either. The hope is dashed, and we abandon for now the idea of 
using the $(n^R - \tau n^{NS}) \cdot f(\zeta)$ production 
structure in Type IIB Calabi--Yau orientifolds,  
until we will recycle this idea in section \ref{ssec:solution}.

\subsection{Deformation around Approximately Arithmetic Complex Structure}
\label{ssec:alg-nmb}

It is easy to imagine that the constraint in 
section \ref{ssec:phen-limit-destabil} is so tight without an idea like the 
one we pursued in section \ref{ssec:first}, that only very negligible 
e-fold is achieved. This statement is so obvious intuitively that we do 
not think it is necessary to build up precise estimate of the e-fold. 
The constraint in section \ref{ssec:phen-limit-destabil} was derived, 
however, by using the potential $V_{\rm racetrack}(T)$ in combination with 
guess work. The technical presentation in the following is the first step 
one needs to take in order to verify or falsify the ``guess work'' part of 
the discussion in section \ref{ssec:phen-limit-destabil}.
We also believe that the following discussion (making an estimate of e-fold)
will remain useful also when some ideas (like those in 
section \ref{ssec:solution}) are implemented.

Obviously we need to take on both of these problems:
\begin{itemize}
\item [(a)] find a vacuum in string theory where the Kallosh--Linde 
tuning (\ref{eq:cond-KL-min}) is achieved at least approximately 
at the potential minimum $(T, z) = (\vev{T}, \vev{z})$, and
\item [(b)] make sure that the volume destabilization constraint is satisfied 
during the inflation / reheating process.
\end{itemize}
A lesson from the study in section \ref{sssec:KYY-IIB} is that the 
issue (a) itself is not an easy problem, because of the quantization 
condition of the flux and hierarchically small value 
we expect for $\vev{W^{(T)}}$.

Let us study the issue (a) a little more systematically than in 
section \ref{sssec:KYY-IIB}. We assume that the vacuum is found in 
the large complex structure ${\rm Im}(\zeta) \gtrsim 1$ region,\footnote{
It is logically possible that inflation takes place in the 
${\rm Im}(\zeta) \gg 1$ region, but the vacuum is in the region 
${\rm Im}(\vev{\zeta}) < 1$. Such a case is not covered 
in this section \ref{ssec:alg-nmb}.} and employ a presentation where 
there is only one complex structure modulus field $\zeta$ (as in the 
mirror quintic); generalization to cases with $h^{2,1}(X) > 1$ may be possible, 
but we will not discuss. 

We consider it is an important clue in thinking about the issue (a) 
that the vacuum value $\vev{W^{(T)}}$ is hierarchically small relatively 
to $M_{\rm Pl}^3/\sqrt{4\pi}$ in the moderately large radius regime (${\rm Re}(\vev{T} \approx {\cal O}(10)$ or a 
little more); the factor $1 \gg 1/[(4\pi)({\rm Re}\vev{T})^{3/2}]$ and 
additional exponential factors are unavoidable. Compared against this 
is the combination $\sum_a (n_a^{R} - n_a^{NS} \tau) \Pi_a$, which does not 
contain any small value. Thus, the tuning condition of Kallosh--Linde 
(\ref{eq:cond-KL-min}) is understood primarily as $\vev{W_{GVW}} \approx 0$
in the sense that $|\vev{W_{GVW}}|$ is much smaller than 
$M_{\rm Pl}^3 \times {\cal O}(1)$. Minimum finding problem of the 
superpotential $W_{GVW}$ with the condition $\vev{W_{GVW}}\approx 0$ has been 
addressed in the literature such as \cite{GVW-zero}.

In the large complex structure region ${\rm Im}(\zeta) \gtrsim 1$ of the 
moduli space, the period integral (\ref{eq:period-integral}) is already 
given in the form of power-series expansion in $e^{2\pi i \zeta}$ 
(world-sheet instanton expansion in the mirror of $X$), and we can 
think\footnote{See footnote \ref{fn:2ways-arithmetic}.} of 
using this approximation scheme in order to give more precise meaning in 
the statement $\vev{W_{GVW}}\simeq 0$ above. 

To be more explicit, we drop all the 
${\cal O}(e^{2\pi i n \zeta})_{n\geq 1}$ parts and also the $i \zeta_3$ term 
from period integrals (\ref{eq:period-integral}) to define 
$\Pi_a^{poly}(\zeta)$, and 
\begin{equation}
  f_R^{poly}(\zeta) := \sum_a n_a^R \Pi_a^{poly}(\zeta), \qquad 
  f_N^{poly}(\zeta) := \sum_a n_a^{NS} \Pi_a^{poly}(\zeta).
\end{equation}
At this level of approximation, we have $W_{GVW}^{poly} = M_{\rm Pl}^3/\sqrt{4\pi}
 \times (f^{poly}_R(\zeta) - \tau f^{poly}_N(\zeta))$. The argument above 
implies that we should require 
\begin{equation}
 [W_{GVW}^{poly}] = 0
\label{eq:KL-tuning-poly-approx-a}
\end{equation}
at the approximate vacuum value $(\tau, \zeta) = (\tau_*, \zeta_*)$
characterized by  
\begin{equation}
\partial_\tau [W_{GVW}^{poly}] = 0, \qquad 
\partial_\zeta [W_{GVW}^{poly}] = 0. 
\end{equation}
This means that 
\begin{equation}
 f^{poly}_N(\zeta_*) = 0, \qquad 
  \tau_* = \frac{\partial_\zeta f^{poly}_R(\zeta_*)}
                {\partial_\zeta f_N^{poly}(\zeta_*)}, \qquad 
f^{poly}_R(\zeta_*) = 0. 
\label{eq:cond-KL-poly}
\end{equation}

With the definition of $f^{poly}_N$ and $f^{poly}_R$ introduced above, 
we can still maintain connection between arithmetics and flux vacua 
 over the entire region of ${\rm Im}(\zeta) \gtrsim 1$, although the connection is now under an 
approximation scheme (see \cite{Moore}).\footnote{\label{fn:2ways-arithmetic}  
The argument here is only to present an idea of how to achieve 
the Kallosh--Linde tuning, and we do not mean to say this 
is the only possibility. Indeed, the vacuum value $\vev{W_{GVW}}$ needs  
to be zero only approximately (relatively to $M_{\rm Pl}^3 \times {\cal O}(1)$).
The relation (\ref{eq:KL-tuning-poly-approx-a}) or $f^{poly}_R(\zeta_*)=0$ 
needs to hold only approximately. An alternative to the idea we 
adopted in the main text is that $\Pi_a(\zeta)$'s and 
$\partial_\zeta \Pi_a(\zeta)$'s take their values in some algebraic 
extension field $K_{(\zeta)}$ and somehow $f_N(\zeta)=f_R(\zeta)=0$ ---($\star$),
rather than $\Pi^{poly}_a(\zeta)$, $\partial_\zeta \Pi_a^{poly}(\zeta)$ do \cite{Moore,DeWolfe}.
In this alternative, $f_N^{poly}=0$ and $f_R^{poly}=0$ may not have a 
common solution, though $\zeta_*$ for $f_N^{poly}(\zeta_*)=0$ should be very close to one of the solutions to $f_R^{poly}=0$.
The true vacuum value of $\zeta$ will be shifted from 
the one satisfying ($\star$), because of the $\zeta$ dependence 
of the prefactors $a_i$ in the gaugino condensation terms. This shift in 
$\vev{\zeta}$, and hence that in $W_{GVW}$ will be exponentially small, 
so that there is still a hope that there is some non-trivial cancellation mechanism 
between this shift and the value $\vev{W^{(T)}}$.} 
In Calabi--Yau orientifolds of Type IIB string, 
$f^{poly}_{N,R}$ are always at most cubic polynomial of $\zeta$, and 
the coefficients take values in $\Q$. The approximate vacuum value 
$(\zeta_*, \tau_*)$ satisfying (\ref{eq:cond-KL-poly}) are always 
algebraic numbers, and they generate an algebraic number field 
$\Q[\zeta_*, \tau_*]$, once all the flux quanta are fixed. 

There are two distinct cases whose consequences are quite different. 
One is the case all the three roots of $f^{poly}_R$ are the same as 
those of $f^{poly}_N$. There is a common $f^{poly}(\zeta)$ so that 
$f^{poly}_R=n_4^R f^{poly}(\zeta)$ and $f^{poly}_N=n_4^{NS} f^{poly}(\zeta)$, 
and we go back to the case we have already seen 
in (\ref{eq:KYY-lookalike-IIB-eqv}).
All the other cases have the property that 
\begin{equation}
 {\rm dim}_\Q \left( \Q[\zeta_*, \tau_*] \right) = 2;
\end{equation}
to see this, it is enough to see that $\zeta_*$ is a root of a 
quadratic polynomial 
\begin{equation}
f^{poly}_N(\zeta)/n_4^{NS} - f_R^{poly}(\zeta)/n_4^{R} \neq 0
\end{equation}
with all the coefficients being rational. 

In the latter case, the two polynomials can be written down as 
\begin{eqnarray}
f_R^{poly}(\zeta) & = & - \frac{5}{6} \; n_4^R 
    (\zeta - \zeta_{*})(\zeta-\bar{\zeta}_{*})(\zeta - \zeta_R), \\
f_N^{poly}(\zeta) & = & -\frac{5}{6} \; n_4^{NS} 
    (\zeta - \zeta_{*})(\zeta-\bar{\zeta}_{*})(\zeta - \zeta_N), 
\end{eqnarray}
with $\zeta_R$ and $\zeta_N$ in $\Q$, and $\zeta_N \neq \zeta_R$.
It follows that 
\begin{equation}
  \tau_* = \frac{n_4^R}{n_4^{NS}} \frac{\zeta_{*}-\zeta_R}{\zeta_{*}-\zeta_N}.
\end{equation}
If $n_4^R \gg n_4^{NS}$, the vacuum value of dilation is in the weak coupling 
region, ${\rm Im}(\tau_*) \gg 1$, and the existence of such a vacuum 
is reliable. 

All the argument, especially with the vacuum value $(\zeta_*, \tau_*)$ 
satisfying (\ref{eq:cond-KL-poly}), does not guarantee that 
$\vev{W_{GVW}}/M_{\rm Pl}^3 \ll {\cal O}(1)$, however. This is because 
we have set aside the $i \chi(X) \zeta_3$ terms from $\Pi^{poly}_a$'s, 
so that we can maintain the connection with arithmetic vacuum value of the 
complex structure moduli/period integrals. In the presence of these terms in the period integral 
$\Pi_a$, in fact, it follows that $\vev{W_{GVW}} \gg M_{\rm Pl}^3$. 
To see this, let us first define $\tilde{f}_R^{poly}(\zeta)$ and 
$\tilde{f}_N^{poly}(\zeta)$, just like $f_R^{poly}(\zeta)$ and 
$f_N^{poly}(\zeta)$, but without dropping the term 
$i\zeta_3 \times \chi(X)/(2\pi)^3$ from the period integral 
in (\ref{eq:period-integral}). 
The solution $\zeta_*$ of $f_N^{poly}(\zeta)=0$ is slightly shifted to be 
$\tilde{\zeta}_*$ for $\tilde{f}^{poly}_N(\tilde{\zeta}_*)=0$. The shift $\tilde{\zeta}_* - \zeta_*$ remains very small, provided $\zeta_*$ 
is in the ${\rm Im}\zeta_* \gg 1$ region and the approximation scheme 
works well. 
For this small shift, however, the (approximate) vacuum value of 
\begin{equation}
 (\tilde{f}_R^{poly}(\zeta)-\tau\tilde{f}^{poly}_N(\zeta))|_{\zeta= \tilde{\zeta}_*}
  = 
 - \frac{5}{6}n_4^R (\zeta_N- \zeta_R) (\zeta - \zeta_*)(\zeta - \bar{\zeta}_*)
  |_{\zeta = \tilde{\zeta}_*},
\end{equation}
is evaluated to be approximately 
\begin{equation}
  - n_4^{NS} {\rm Im}(\tau_*) \times 
   i n_4^{NS} \frac{\chi(X) \zeta_3}{(2\pi)^3};
\end{equation}
this means that we need to give up either ${\rm Im}(\tau_*)\gg 1$ or 
$\vev{W_{GVW}} \ll M_{\rm Pl}^3$. 

The discussion above also reveals that there is an obvious loop hole.
Calabi--Yau 3-folds with $\chi(X)=0$ have a special properties that 
the $i \zeta_3/(2\pi)^3$ term drops out from the period integral (see \cite{Moore} however).  
In this case, $\tilde{f}_N^{poly}=f_N^{poly}$ and $\tilde{f}_R^{poly} = f_R^{poly}$, 
so that $\zeta = \zeta_*$ remains to be the solution of both. The 
Kallosh--Linde tuning condition is maintained at least at the level of terms 
of order $M_{\rm Pl}^3 \times {\cal O}(1)$; cancellation involving the 
``mirror-worldsheet-instanton'' correction terms ${\cal O}(e^{2\pi i n \zeta})$'s 
and $\vev{W^{(T)}}$ remains to be an open problem, however. The string coupling 
can be small at the vacuum, because 
\begin{equation}
{\rm Im}(\tau_*) = \frac{n^R_4}{n^{NS}_4}
  {\rm Im}\left[ \frac{\zeta_N-\zeta_R}{\zeta_*-\zeta_N} \right]
\end{equation}
can be made much larger than unity by taking $(n^R_4/n^{NS}_4)$ sufficiently 
large. Unlike in the cases we studied in section \ref{sssec:KYY-IIB}, 
we have a vacuum in the reliable weak coupling regime. 
This holds as an approximate solution\footnote{To be rigorous, one should also ask whether there is a D7-brane configuration for the $SU(N_1)\times SU(N_2)$ gaugino condensation satisfying the calibration condition in the presence of fluxes. F-theory will be a better tool to study this issue, however.} to the issue (a).

With a vacuum of complex structure formulated, it is now possible to discuss 
how much the complex structure can be deformed from the vacuum value during 
the slow-roll inflation/reheating process, i.e., the issue (b). 
Since we have imposed conditions that $W_{GVW} \approx 0$ and 
$\partial_\zeta W_{GVW} \approx 0$ at $\zeta \approx \zeta_*$, 
the value in $\delta \widetilde{W}^{({\rm cpx})}_{\rm eff.}$ can be evaluated by using 
\begin{equation}
\left. \frac{\partial^2 }{\partial \zeta^2} 
   \left[f_R^{poly}(\zeta)-\tau_* f_N^{poly}(\zeta)\right] \right|_{\zeta = \zeta_*}
   = - \frac{5}{3} n_4^R (\zeta_* - \bar{\zeta}_*)
     \frac{\zeta_R-\zeta_N}{\zeta_*-\zeta_N} \sim 
  {\rm Im}(\tau_*) \times n_4^{NS} (\zeta_* - \bar{\zeta}_*).
\end{equation}
We have chosen parameters, so that this second derivative is larger than 
${\cal O}(1)$; ${\rm Im}(\zeta_*) \gg 1$ so that the approximation of 
dropping the mirror world-sheet instanton terms are small, and the string 
coupling is also small (${\rm Im}(\tau_*) \gg 1$). The constraint 
(\ref{eq:Wtil-cpx-eff-limit}) therefore implies that the complex structure 
deformation $\delta \zeta = \zeta - \zeta_*$ is allowed without volume 
destabilization, at most only at the level of $\sqrt{10^{-3}/{\rm Im}(\zeta)}$.
This is much smaller than the variation 
$\delta \zeta \sim {\rm Im}(\zeta) \times {\cal O}(10-100)$ we need for 
sufficient e-fold.  

\subsection{Possible Solutions of Inflation}
\label{ssec:solution}

We have seen so far that it is not easy at all for complex structure moduli 
field to play some role in inflation.  If we still try to seek for 
the ``right-handed sneutrino scenario'' to be relevant somehow in the 
Type IIB/F-theory compactifications with moderately large internal volume, 
that will have to be in a very special situation. The discussion in this 
article helps us pin down such possibilities. 
Let us describe a few of these at the end of this article. 

{\bf I}: The first possibility is still to try to seek for an analogue 
of \cite{KYY} or (\ref{eq:KYY-lookalike-IIB}), so that 
$(\delta \zeta)^2 \times (\partial^2_\zeta W_{GVW})|_{\zeta = \zeta_*}$ remains 
small or vanishes during inflation. At least two fields are necessary then, 
and some structure needs to be present in the space of complex structure 
moduli. Although we have seen in section \ref{sssec:KYY-IIB} that 
the flux superpotential for Type IIB orientifolds (\ref{eq:KYY-lookalike-IIB})
does not work, we can think of F-theory compactification over a Calabi--Yau 
4-fold $Y_4 = {\rm K3} \times {\rm K3} = S_1 \times S_2$, where the 
period integral is in a factorized form 
$\Omega_{Y} = \Omega_{S_1} \cdot \Omega_{S_2}$. This is an analogue 
of (\ref{eq:KYY-lookalike-IIB}) in Type IIB orientifolds, 
in that the factorized form of $(n^R - \tau n^{NS}) \cdot \Pi$ 
originates from the product structure in $Y_4 = (T^2\times X)/\Z_2$. 
This F-theory version also has advantages of being able to handle 
large $g_s$ region of the moduli space, and of being able to generate 
up-type Yukawa couplings in supersymmetric unified theories through 
its $E_6$ algebra \cite{F4yukawa}.
It is known in the $Y_4 = {\rm K3} \times {\rm K3}$ compactification of 
F-theory that some type of flux [a part of what is called ``$G_0$-type'' 
in \cite{AK}] preserves the factorized structure of the period integral and 
gives rise to pure Dirac type mass matrix in the superpotential \cite{BKW}. 

A down side of the K3 x K3 compactification of F-theory is that all 
the 7-branes are parallel, so that quarks and leptons are not generated 
in 7-brane intersections. Thus, instead of taking a Calabi--Yau 4-fold to 
be purely in a direct product, $Y_4 = S_1 \times S_2$, one may think of 
$Y_4$ being a K3-fibration\footnote{It follows in this case \cite{BHV-2} 
that an elegant mechanism of 
$\SU(5)_{\rm GUT} \longrightarrow \SU(3)_C \times \SU(2)_L \times \U(1)_Y$ 
symmetry breaking in \cite{Buicanetal, BHV-2} cannot be used. There are 
other mechanism for this symmetry breaking, however; see \cite{BHV-2, TW-GUT} 
for more information.} over a complex surface $S$, and hope that something 
similar to the product structure of the period integral 
$\Pi_{S_1} \times \Pi_{S_2}$ still remains even in the fibred geometry, 
at least approximately somewhere in the moduli space. The authors 
do not know whether that is true. 
References \cite{GrimmKlemm, Klemm} contain useful information 
in extending the study of section \ref{ssec:alg-nmb} 
in such F-theory compactifications.\footnote{The analogue of $\chi(Z)=0$ 
Calabi--Yau 3-folds in section \ref{ssec:alg-nmb} will be a 4-fold $Y_4$ 
whose mirror $\widetilde{Y}_4$ has vanishing third Chern class.}

{\bf II}: As a second possibility, one can think of topological inflation, 
if the $W_{GVW}$ superpotential for a given flux admits multiple minima. 
One minimum may be in the large complex structure region, 
${\rm Im}\zeta \gtrsim 1$, though the other minimum may not be.
It is a rule of thumb that topological inflation takes place if the 
distance (measured in the metric of the non-linear sigma model) between 
the minima exceeds $M_{\rm Pl}$. At least by choosing flux quanta 
properly,\footnote{We can think of flux of the following property: 
$n^{NS}_4 \approx {\cal O}(1)$, $n^{NS}_3 \approx O(K)$, 
$n^{NS}_2 \approx O(K^2)$ and $n^{NS}_1 \approx O(K^3)$, 
where $K$ is a number 
much larger than 1. When the NS-NS 3-form flux is chosen in that way, 
all of the three solutions to $f^{poly}_N(\zeta)=0$ satisfy 
$|\zeta| \approx O(K)$, one of them has an ${\cal O}(K)$ positive imaginary part, another has an 
${\cal O}(K)$ negative imaginary part, and 
the last one, $\zeta_N$, being real. Thus, one of the three solutions, 
(and in fact only one of them), denoted by $\zeta_*$, is in the region 
where ${\rm Im}\zeta \gg 1$. Similar argument holds also for the roots of 
$f_R^{poly}=0$. It should be remembered that we cannot take this scaling parameter $K$ to be arbitrarily large, however. There is an upper bound in the flux contribution to the D3-tadpole. 
$\Im (\zeta _*)$ should not be too large either, because winding stringy states will be in the light spectrum \cite{Oogurivafa}.}
one of the minima can be moved to large ${\rm Im}\zeta$ region. 
In the middle of a domain wall in between the two minima, the volume 
stabilization constraint discussed in section \ref{sec:barrier} 
may no longer be satisfied, however. 
The authors are not sure whether such a field configuration leads to 
topological inflation, and whether the predicted fluctuations in CMB are 
consistent with all the observations.

{\bf III}: One cannot rule out logically that the complex structure $\zeta$ 
dependence of the prefactors $a_i$ in the racetrack superpotential is such 
that the barrier in the volume stabilization potential remains high enough 
under the evolution of $\zeta$ during inflation 
(c.f. \cite{HKKLL}).

{\bf IV}: The bottom-up idea of ``right-handed sneutrino inflation'' 
scenario has two ingredients: i) the scalar partner of the right-handed 
neutrinos and the Majorana mass provides the potential of chaotic inflation, 
and ii) the reheating process proceeds dominantly through the renormalizable 
interactions of right-handed neutrinos. Even when inflation is 
driven  purely by the K\"{a}hler moduli field $T$, the reheating process 
may proceed through right-handed neutrinos (which are part of 
complex structure moduli \cite{RHnu-Ftheory}), due to the mixing between 
the K\"{a}hler moduli and complex structure moduli fields, as we discussed 
in section \ref{sec:KL-problem}. Therefore, the volume destabilization limit 
on the complex structure deformation can be avoided, but the aspect ii) of 
the right-handed sneutrino inflation scenario is still realized in this case.
Various predictions on the thermal history after inflation will be lost, 
however, because the reheating process depends on various details of 
spectrum and mixing of K\"{a}hler and complex structure moduli at the vacuum, 
not just on the physics of right-handed neutrinos.

\subsection*{Acknowledgements}

We thank T.T.Yanagida for stimulating discussion, and for giving us his 
comments on our manuscript. 
HH would like to thank Mainz Institute for Theoretical Physics for hospitality and its partial support during a part of this work.
This work is supported by the REA grant agreement PCIG10-GA-2011-304023 from the People Programme of FP7 (Marie Curie Action) (H.H.), the grant FPA2012-32828 from the MINECO (H.H.), the ERC Advanced Grant SPLE under contract ERC-2012-ADG-20120216-320421 (H.H.), the grant SEV-2012-0249 of the ``Centro de Excelencia Severo Ochoa" Programme (H.H.), 
WPI Initiative (R.M., T.W.), Advanced Leading Graduate Course
for Photon Science grant (R.M.), and a Grant-in-Aid for Scientific Research on 
Innovative Areas 2303, MEXT, Japan (T.W.).

\end{document}